\documentclass[a4paper,10pt]{article}
\usepackage{amssymb}
\usepackage{txfonts}
\usepackage{subfigure}
\usepackage{booktabs}
\usepackage{eurosym}
\usepackage[pdftex]{graphicx}

\usepackage{multirow}
\usepackage{booktabs}
\usepackage{graphicx}
\usepackage{amssymb}
\usepackage{eepic}
\usepackage{amscd}
\usepackage{longtable}
\usepackage{array}

\newcommand{\nc}{\newcommand}
\nc{\rnc}{\renewcommand}
\nc{\nn}{\nonumber}

\newcommand{\sss}{\scriptscriptstyle}

\title{Density profiles of the exclusive queueing process}

\author{
Chikashi Arita\thanks{
Institut de Physique Th\'{e}orique, CEA Saclay. 
chikashi.arita@cea.fr}
and
Andreas Schadschneider\thanks{
Institut f\"{u}r Theoretische Physik, Universit\"{a}t zu K\"{o}ln.
as@thp.uni-koeln.de}
}

\date{}

\begin{document}

\maketitle

\begin{abstract}
The exclusive queueing process (EQP) incorporates the exclusion
principle into classic queueing models. It can be interpreted
as an  exclusion process of variable system length.
Here we extend previous studies of its phase diagram by identifying
subphases which can be distinguished by the number of plateaus 
in the density profiles. Furthermore the influence of different 
update procedures (parallel, backward-ordered, continuous time)
is determined.
\end{abstract}


\section{Introduction}\label{sec:intro}

Queueing theory is one of the most important topics in the field of
operations research \cite{ref:Medhi,ref:Saaty,ref:factory}.  It has a
broad spectrum of applications ranging from telecommunications to
traffic engineering and supply chains. One of the simplest queueing
processes is the so-called M/M/1 model, where customers enter the
system with probability $\alpha$ and leave the system with probability
$\beta$ at one server.  The current state of the M/M/1 queueing
process is completely specified by the number of customers.  The
system converges to a stationary state with a finite number 
of customers when $\alpha<\beta$ whereas the number of waiting
customers diverges for $\alpha > \beta$.

A feature which seems to be important for pedestrian queues and other
traffic applications is the excluded-volume effect: pedestrians can
proceed only when there is enough space in front of them
\cite{ref:SCN}.  This is e.g.\ seen in queues at the check-in at
airports where passengers have to move the luggage when moving
forward.  However, standard queueing models like the M/M/1 model
neglect the excluded-volume effect, and do not have a spatial
structure.  Then  the length $L$ of the system is given by the number of
waiting customers $N$ (if customers have unit length) and the density
is constant in space ($\rho=N/L=1$).  

\begin{figure}
\begin{center}
\centering
\includegraphics[width=0.5\columnwidth]{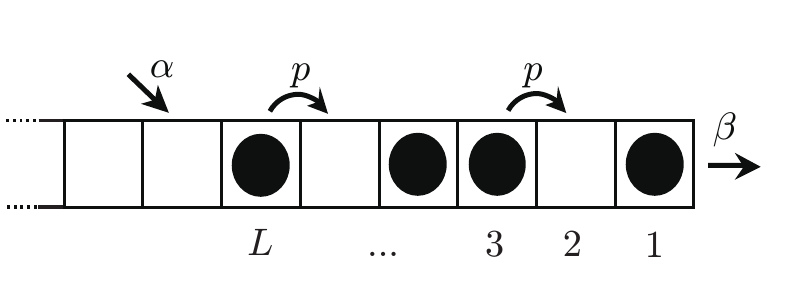}
\caption{Exclusive queueing process.}
\label{fig:eq} 
\end{center}
\end{figure}

The ``exclusive queueing process'' (EQP) was introduced in
\cite{ref:A,ref:Y} to investigate how the excluded volume affects
queues.  It is obtained by modifying the input procedure of the
one-dimensional totally asymmetric simple exclusion process (TASEP)
with the ordinary ``open boundary condition.''  Customers are injected
always at the end of the queue, called left end in the following,
i.e.\ behind the last customer waiting (Fig.~\ref{fig:eq}). This is in
contrast to the usual open TASEP where the input always happens at the
same site, irrespective of the occupation of the other sites.  The
output is not changed compared to the usual open TASEP: in both
models customers are extracted at the right end (the server) which is
fixed.

The EQP is not the only variant of the TASEP on a dynamic lattice.
An earlier example is the dynamically extending
exclusion process (DEEP) introduced in
 \cite{Evans1,Evans2,Evans3,ref:DMP}
as a model for fungal growth.
 In contrast to the EQP, the DEEP
has no mechanism for reducing the system length and therefore
the length of the system is always diverging. Other possible biological applications are length-regulation of microtubules
\cite{ref:MRF} and bacterial flagellar growth \cite{ref:SchS}.

The state space of the EQP is the set of configurations of the
customers. It is more precisely given as
\begin{eqnarray}
  S=    \{ \tau = 1\tau_{L-1}\cdots \tau_1
    | L\in \mathbb Z_{\ge 0},  \tau_j=0,1 \} ,
\end{eqnarray}
where $\tau= \emptyset,1$ for $L=0,1$, respectively.
The state $\tau=1\tau_{L-1} \cdots \tau_1 \in S\setminus\{\emptyset\}$
corresponds to the customer configuration where each site $j$ is
occupied or empty according to $\tau_j=1$ or 0, and
$L$ defines the length of the system.
The symbol $\emptyset$ corresponds to the state ``no
customer in the system.'' 
  We denote
the number of customers ($\#(\tau_j=1) $) by $N$ for a given state $
\tau=1\tau_{L-1} \cdots \tau_1$.  Each customer enters the system,
hops and leaves the system as
\begin{eqnarray}
\begin{array}{rcrl}
  \emptyset  & \to &  1 &
     {\rm with\  probability\ } \alpha, \\
  \ \, 1 \cdots  & \to &  11 \cdots &
     {\rm with\  probability\ } \alpha, \\
  \cdots 10 \cdots  & \to & \cdots  01 \cdots &
     {\rm with\  probability\ } p, \\
  \cdots 1  & \to &  \cdots  0  &
     {\rm with\  probability\ } \beta .
\end{array}
\label{eq-localrules}
\end{eqnarray}
The local update rules (\ref{eq-localrules}) are not sufficient to
specify the dynamics fully.  In addition the sequence in which the
rules are applied to the sites or particles needs to be defined. Here
we consider two discrete-time updates
(parallel and backward sequential updates)
 and a continuous-time dynamics where the parameters
$\alpha$, $\beta$ and $p$ are transition rates (not probabilities).

In \cite{ref:A,ref:AY}, exact stationary states for the
continuous-time case and the parallel-update case were constructed as
matrix product states based on the known forms for the corresponding
TASEPs with a fixed length \cite{ref:DEHP,ref:ERS}.
The phase boundary between the convergent and divergent phases was
found to be modified compared to the classical M/M/1 queue.  
In particular, for the convergence to the stationary states, the
injection rate (or probability) $\alpha$ cannot be bigger than the
maximal current of the TASEP, i.e. ``the queue itself is a
bottleneck'' as well as the server.

In \cite{ref:AS1}, the phase diagram was analyzed in more detail.  The
convergent and divergent phases are both further subdivided in two
subphases analogous to the maximal current and high-density phases of
the TASEP.  Furthermore time dependent properties were investigated.
However the asymptotic form of the velocity for the growth
of $L$ in the divergent phase was left as an open problem.  In this
article we will clarify this point, giving density profiles with help
of Monte Carlo simulations.
(In the case where the customer hopping is
deterministic $(p=1)$, an exact ``dynamical state'' in matrix product form
exists which enables us to rigorously derive the behavior of $\langle
L_t\rangle$ and $\langle N_t\rangle$ \cite{ref:AS2}.)

This article is organized as follows.  In
Section~\ref{sec:update-rules}, we define the EQPs with various
updates in more detail.  The phase diagrams are derived and their
relation with classical queueing processes is discussed.  In
Section~\ref{sec:subphases} , based on simulation results we
characterize further subphases of the divergent phases according to
the shapes of the density profiles.  In Section~\ref{sec:critical}, we
investigate the EQPs on the critical line separating the divergent and
convergent phases.  Finally, we give a summary and conclusions of this
article in Section~\ref{sec:conclusion}.


\section{Update rules}\label{sec:update-rules}

The TASEP is a prototypical model of stochastic interacting particle
systems.  It has been studied intensively in the last decades both
from the view of nonequilibrium statistical physics
\cite{ref:D,ref:S,ref:BE,ref:SCN} and mathematics (see e.g.\ 
\cite{ref:L}).  Similar to the TASEP with the ordinary open boundary condition 
 \cite{ref:RSSS} one can study the
EQP with different update schemes.  In the next two subsections we
define two discrete-time EQPs and determine their phase diagrams. These
are divided into four phases according to their asymptotic lengths
(convergent vs.\ divergent) and the parameter-dependence of the
outflows.  Then we consider special cases and limits including
the continuous-time EQP.
\begin{figure}[h]
\begin{center}
\includegraphics[width=0.32\columnwidth]{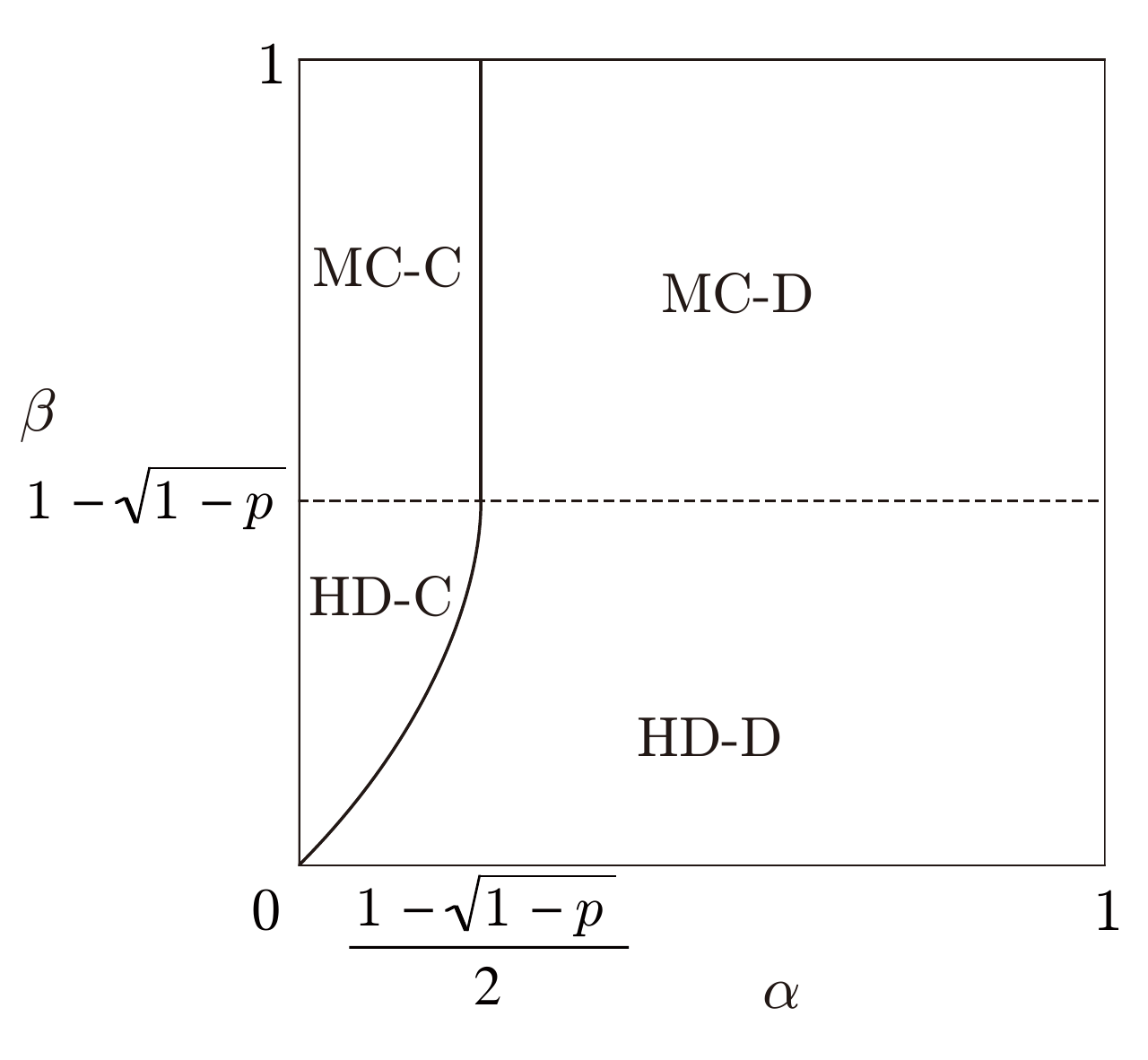}
\includegraphics[width=0.32\columnwidth]{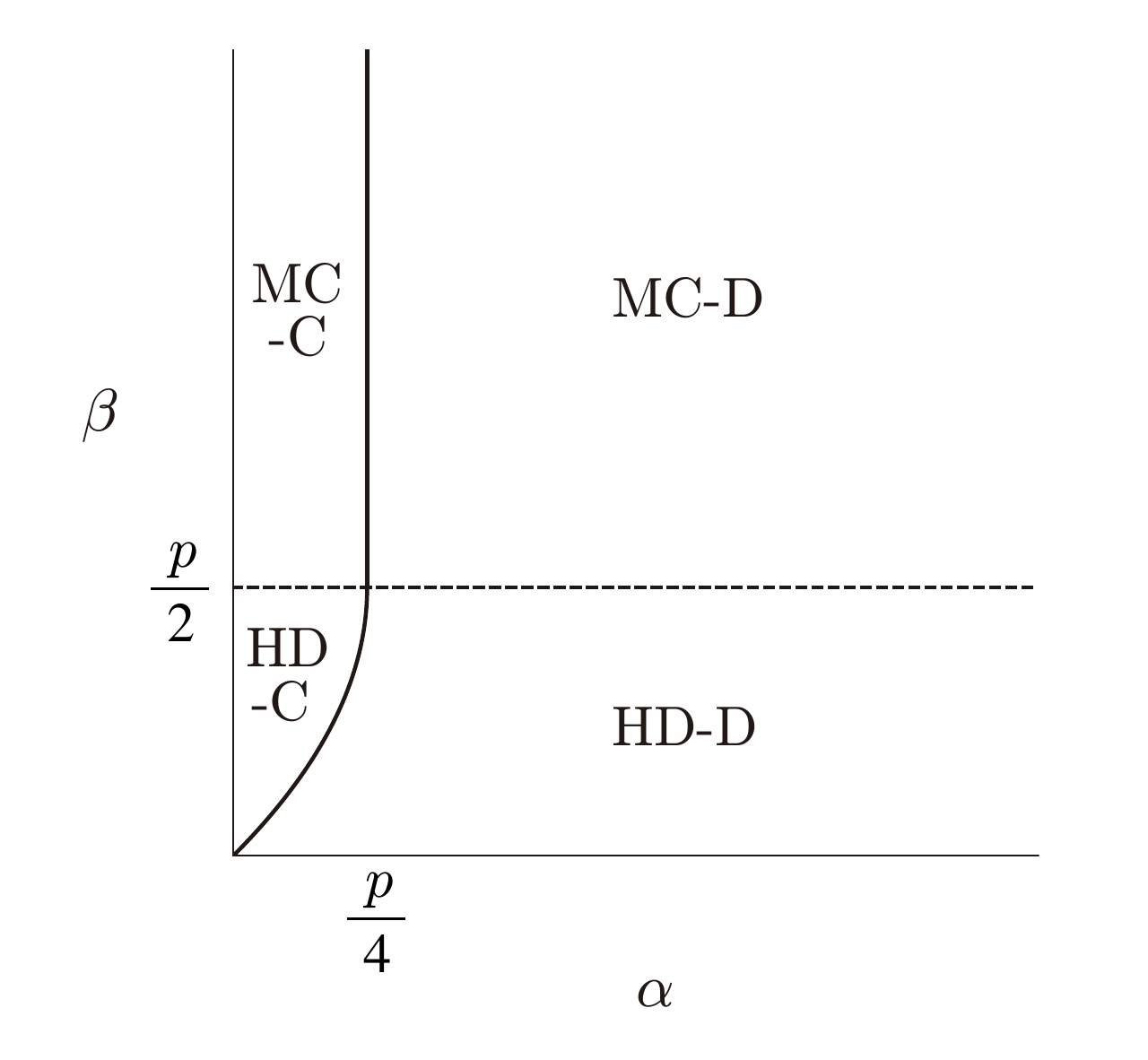}
\includegraphics[width=0.32\columnwidth]{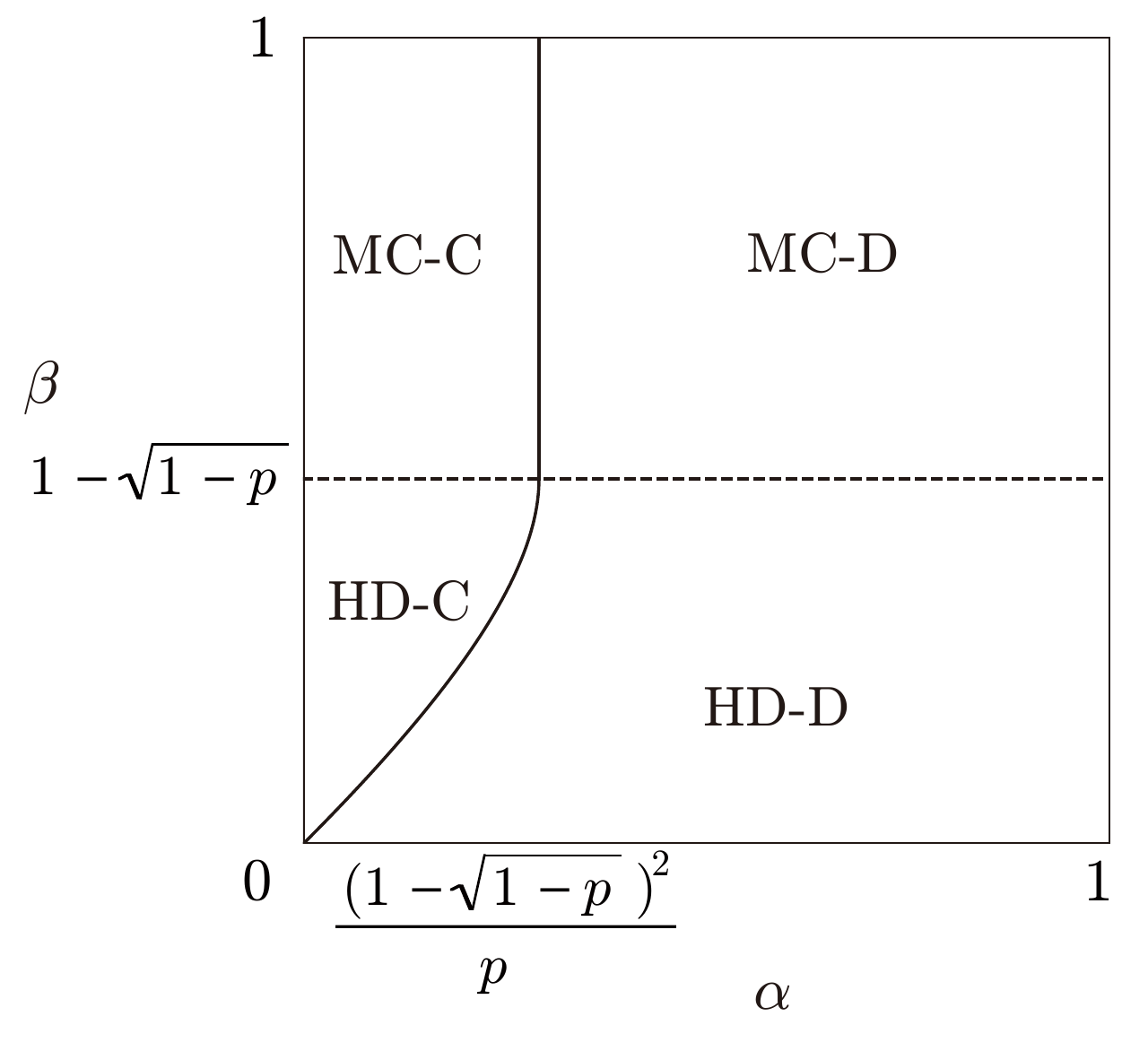}
\\
\includegraphics[width=0.32\columnwidth]{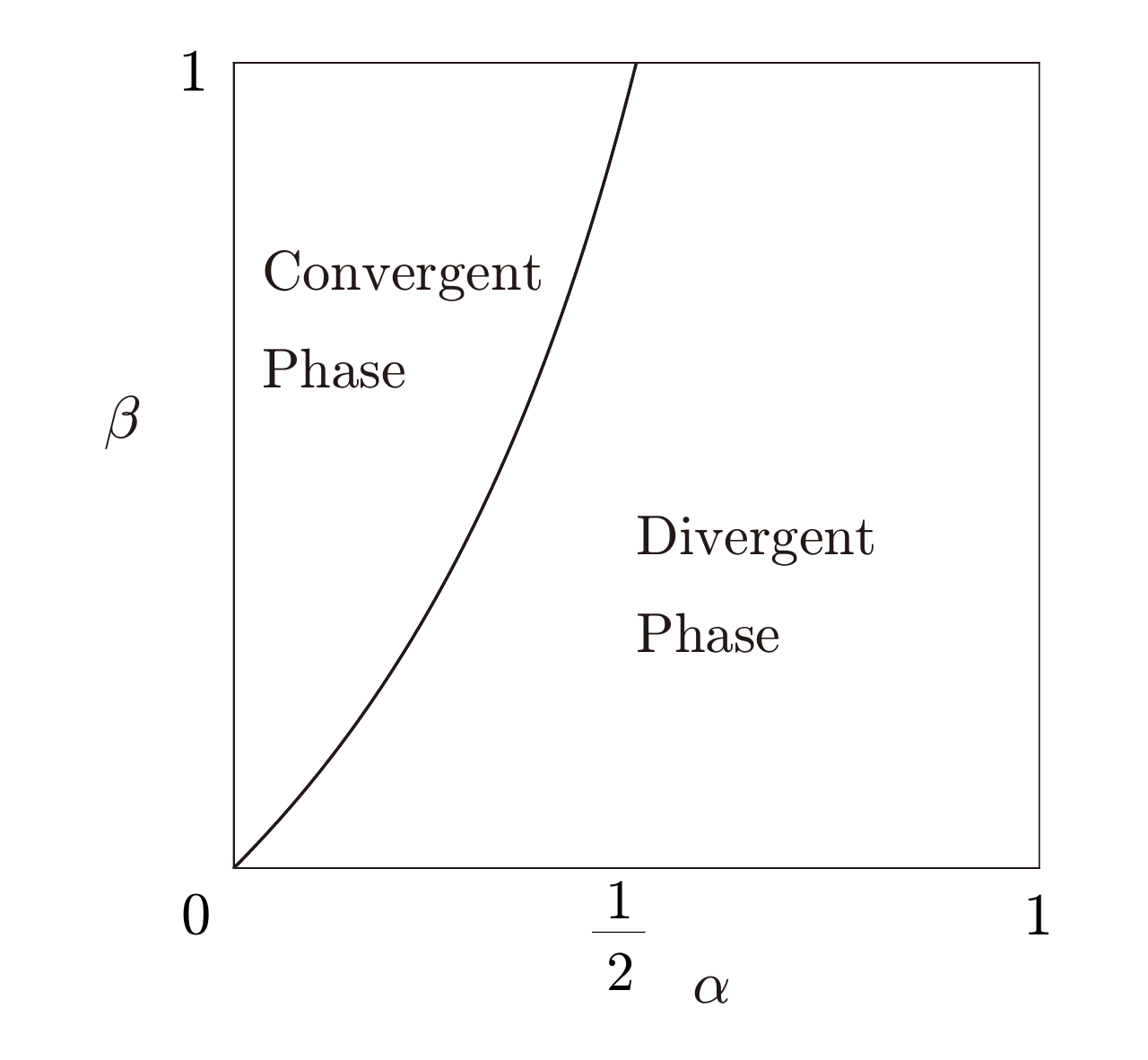}
\includegraphics[width=0.32\columnwidth]{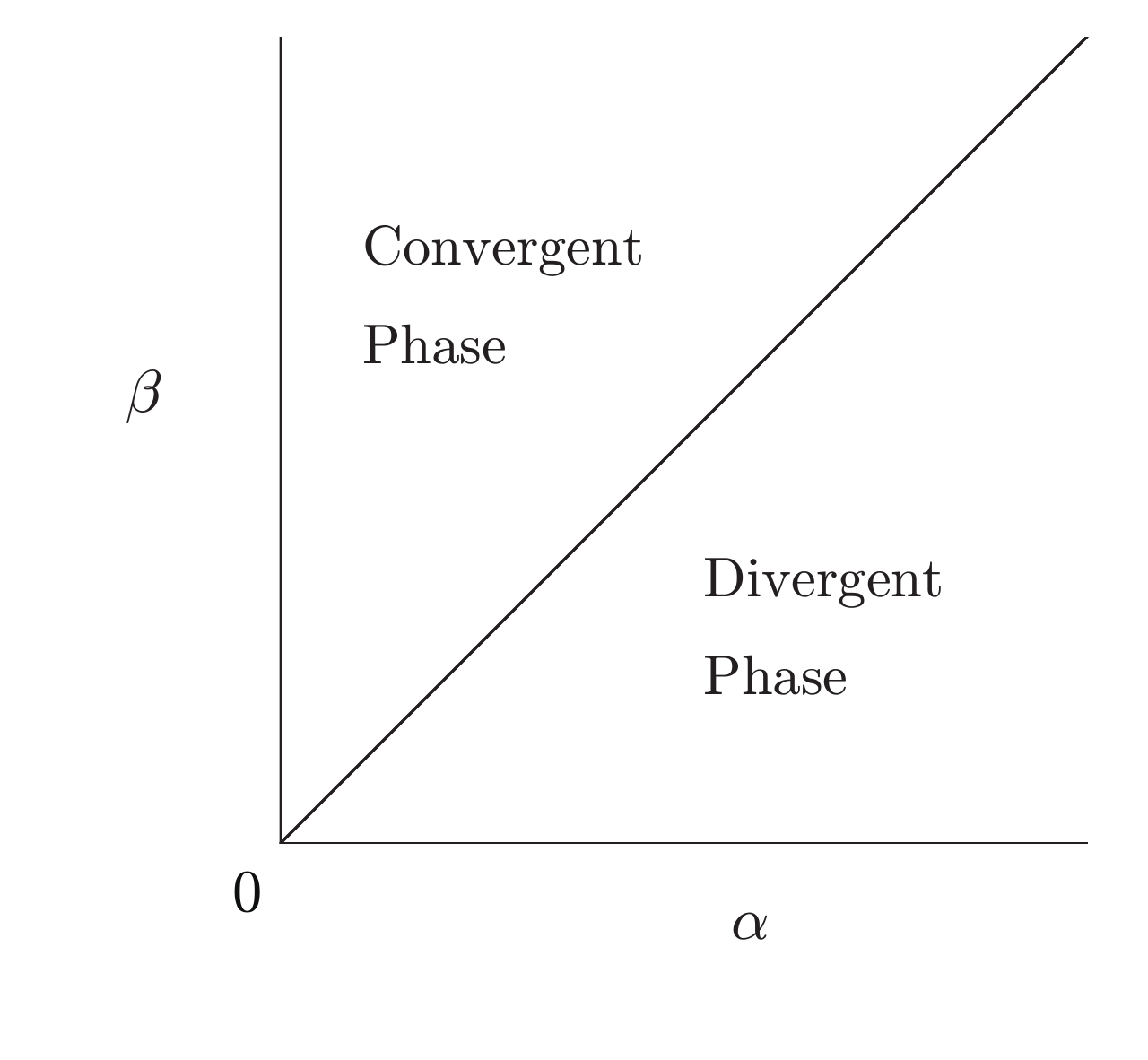}
\includegraphics[width=0.32\columnwidth]{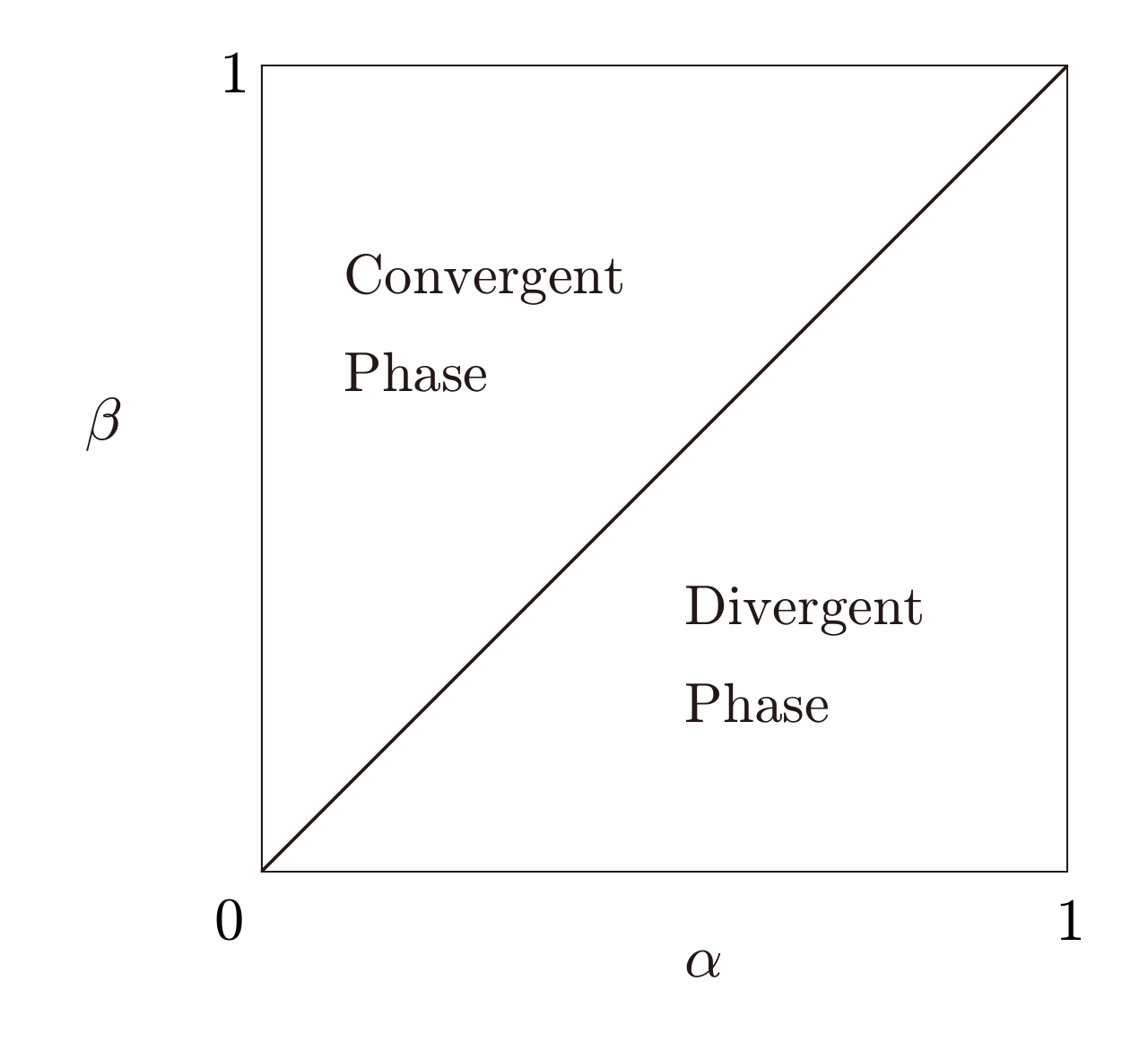}
\vspace{-5mm}
\caption{
  The phase diagrams of various queueing processes: the parallel EQP
  (top-left), the continuous-time EQP (top-middle), the backward EQP
  (top-right), the rule 184 CA with stochastic boundaries
  (bottom-left), the continuous-time M/M/1 (bottom-middle) and the discrete
  M/M/1 (bottom-right).  The relations among the phase diagrams can be
  seen by (\ref{eq:relation-queues}).  For example the bottom-left is
  obtained by setting $p=1$ in the top-left.  }
\label{fig:phase-diagrams}
\end{center}
\end{figure}

\subsection{Parallel update}

In \cite{ref:Y,ref:AY,ref:AS1,ref:AS2}, the EQP with the parallel
update rule, which we call parallel EQP shortly, was investigated.
In the parallel dynamics all sites are updated simultaneously, e.g.
\begin{eqnarray}\label{ref:example-para}
\begin{array}{rcrl}
    101011 & \to &  1011010 &
  {\rm with\  probability}
   \ \alpha\times p\times(1-p)\times\beta,  \\
    111011 & \to & 110111 &
  {\rm with\  probability}
  \ (1-\alpha)\times p\times (1-\beta),  \\
\emptyset & \to & \emptyset &
  {\rm with\  probability}
  \  1-\alpha   .
\end{array}
\end{eqnarray}

The current-density relation for the TASEP with the parallel update 
and the periodic boundary condition  is given by \cite{ref:SS,ref:SSNI}
\begin{eqnarray}
\label{eq:J-rho-para}
J_{\sss \parallel}(\rho)
=\frac{1-\sqrt{1-4p\rho(1-\rho)}  }{2} ,\quad
\end{eqnarray}
which is also true 
for the ordinary open boundary condition \cite{ref:ERS}.
The ``critical line'' that separates the parameter space into 
divergent and convergent phases is given by $ \alpha = J_{\sss
  \parallel}(\rho_{\sss\parallel}) $, where
\begin{eqnarray}
\label{eq:crit-line-para}
J_{\sss \parallel}(\rho_{\sss\parallel}) =
\left\{ \begin{array}{ll}
    \frac{\beta(p-\beta)}{p-\beta^2} 
       & (0<\beta\le \beta_c ) ,\\
       \frac{ 1-\sqrt{1-p} }{2}
       & (\beta_c < \beta \le 1) ,
   \end{array} \right.
\quad {\rm where\ \ }
 \rho_{\sss\parallel}=
\left\{ \begin{array}{ll}
   \frac{p-\beta}{p-\beta^2} & ( 0<\beta\le \beta_c ), \\
   \frac{1}{2} & (\beta_c < \beta \le 1), 
\end{array}  \right.
\end{eqnarray}
with $\beta_c=1-\sqrt{1-p}$.  When $\alpha<
J_{\sss \parallel}(\rho_{\sss\parallel}) $, the system converges to a
stationary state which has a matrix product form \cite{ref:AY}.  The
number of customers $\langle N_t\rangle$ decreases approximately
linearly in time as
\begin{eqnarray}\label{eq:N-conv-para}
   \langle N_t\rangle
   \sim  (\alpha -  J_{\sss \parallel}(\rho_{\sss\parallel}) ) 
    t+\langle N_0\rangle  
\end{eqnarray}
while $ t\lesssim \frac{\langle N_0\rangle }{ J_{\sss\parallel}
  (\rho_{\sss\parallel}) -\alpha }$ starting from a sufficiently large
$\langle N_0\rangle$ at time $t=0$.  The quantity $J_{\sss
  \parallel}(\rho_{\sss\parallel})$ is actually the customer current
through the right end, i.e. the outflow.
The system length exhibits a similar behavior
\begin{eqnarray}
\label{eq:L-conv-para}
   \langle L_t\rangle    \sim \frac{  \alpha -  
   J_{\sss \parallel}(\rho_{\sss\parallel}  )  }{\rho_{\sss\parallel} }t
   +\langle L_0\rangle ,
\end{eqnarray}
where the density profile is almost flat with the bulk density 
$\rho_{\sss\parallel}$.
In view of the form (\ref{eq:crit-line-para}), 
we call the region $\alpha< J_{\sss \parallel}(\rho_{\sss\parallel}) $
with $\beta>\beta_c$ ``maximal-current-convergent (MC-C) phase'',
and $\alpha< J_{\sss \parallel}(\rho_{\sss\parallel}) $
with $\beta< \beta_c$ ``high-density-convergent (HD-C)  phase''.

When $\alpha> J_{\sss \parallel}(\rho_{\sss\parallel}) $, the system
does not have a stationary state, and $ \langle N_t\rangle$ and $
\langle L_t\rangle$ diverge linearly in time.  For $ \langle
N_t\rangle$, the form (\ref{eq:N-conv-para}) is valid and we have the
asymptotic behavior
\begin{eqnarray}
       \langle N_t\rangle
   \simeq  (\alpha -  J_{\sss \parallel}(\rho_{\sss\parallel}) ) t \quad
   (t\to \infty).
\end{eqnarray}
In view of the form (\ref{eq:crit-line-para}), 
we call the region $\alpha> J_{\sss \parallel}(\rho_{\sss\parallel}) $
with $\beta> \beta_c$ ``maximal-current-divergent (MC-D) phase'',
and $\alpha> J_{\sss \parallel}(\rho_{\sss\parallel}) $
with $\beta< \beta_c$ ``high-density-divergent (HD-D)  phase''.
On the other hand, the form (\ref{eq:L-conv-para}) is not always valid
in the divergent phase (see eq.~(\ref{eq:velo-L-para}) below).  The main
purpose of this paper is to determine the velocity $V_{\sss\parallel}$
for $\langle L_t\rangle\simeq V_{\sss\parallel}t$ as well as the
density profile in the divergent phase.

\subsection{Backward sequential update}

We now consider the discrete-time EQP with backward-sequential update
(backward EQP): first a customer arrives with probability $\alpha$,
and the customer at the right end is extracted with probability
$\beta$ (if it exists).  Then starting from the rightmost particle
and going sequentially to the left up to the leftmost particle,
we move each  particle forward with probability $p$  if possible.
For example
\begin{eqnarray}\label{ref:example-back}
\begin{array}{rrrl}
    101011 & \to & 1011001 &
  {\rm with\  probability}
  \ \alpha\times\beta\times p
  \times (1-p) \times p \times (1-p),\qquad\qquad  \\
    111011 & \to & 11111 &
  {\rm with\  probability}
  \ (1-\alpha)\times(1-\beta)  
  \times p \times p \times p,  \\
\emptyset &  \to & \emptyset &
  {\rm with\  probability}
  \ (1-\alpha)   + \alpha\times \beta .
\end{array}
\end{eqnarray}
In the first example of transitions (\ref{ref:example-back}),
the customer on the second site can move to 
the rightmost site, thanks to the backward update.
On the other hand, he/she cannot move in the 
parallel case, see the first example of Equation (\ref{ref:example-para}).

The current-density relation for the backward-sequential update\footnote{Note 
that the sitewise and particlewise ordered
updates \cite{ref:RSSS} are identical here.} TASEP is \cite{ref:RSSS}
\begin{eqnarray}
\label{eq:J-rho-back}
J_{\sss \gets}(\rho)
=\frac{p\rho(1-\rho) }{1-p\rho } .\quad
\end{eqnarray}
Simulation results imply that the critical line separating the
parameter space of the EQP into convergent and divergent phases is
given by
\begin{eqnarray}
  \alpha =  J_{\sss\gets}(\rho_{\sss\gets}) =
\left\{\ \begin{array}{ll}
     \frac{\beta(p-\beta)}{p(1-\beta)}
       & (0<\beta\le \beta_c ) ,\\
       \frac{(1-\sqrt{1-p})^2 }{p}
       & (\beta_c < \beta < 1) ,
   \end{array} \right.
\quad {\rm where\ \ }
 \rho_{\sss\gets}=
\left\{\ \begin{array}{ll}
  \frac{ p-\beta }{ p(1-\beta) } 
  & ( 0<\beta\le \beta_c ), \\
   \frac{1-\sqrt{1-p}}{p}
    & (\beta_c < \beta < 1) ,
\end{array} \right.
\end{eqnarray}
with $\beta_c = 1-\sqrt{1-p}$ as in the parallel case.  The phase
diagram is qualitatively similar to that of the parallel EQP
(\ref{eq:crit-line-para}), see Figure \ref{fig:phase-diagrams}.  In
the special case $\beta=1$ the system always has the stationary state
$P(\emptyset)=1$ and $P({\rm otherwise})=0$.  Therefore we set $0<\beta<1$
in the following.

When $\alpha< J_{\sss \gets}(\rho_{\sss\gets}) $,
we expect the system to converge to a stationary state, and the number
of customers $\langle N_t\rangle$ decreases approximately linearly in
time as
\begin{eqnarray}\label{eq:N-conv-back}
   \langle N_t\rangle
   \sim  (\alpha -  J_{\sss \gets}(\rho_{\sss\gets}) ) t+\langle N_0\rangle  
\end{eqnarray}
while $ t\lesssim \frac{\langle N_0\rangle }{ J_{\sss
    \gets}(\rho_{\sss\gets}) -\alpha }$ starting from a sufficiently
large $\langle N_0\rangle$ at time $t=0$.  The system length also
exhibits a similar behavior
\begin{equation}
\label{eq:L-conv-back}
   \langle L_t\rangle
   \sim \frac{  \alpha -  J_{\sss \gets}(\rho_{\sss\gets})}
    {\rho_{\sss\gets} }t+\langle L_0\rangle .
\end{equation}

When $\alpha> J_{\sss \gets}(\rho_{\sss\gets}) $,
we expect that the system does not
has a stationary state and $ \langle N_t\rangle$ and $ \langle
L_t\rangle$ diverge linearly in time.  For $ \langle N_t\rangle$, the
form (\ref{eq:N-conv-back}) is valid and we have the asymptotic behavior
\begin{equation}
       \langle N_t\rangle
   \simeq  (\alpha -  J_{\sss \gets}(\rho_{\sss\gets}) ) t \quad
   (t\to \infty).
\end{equation}
On the other hand, for the divergence of the length the form
(\ref{eq:L-conv-back}) is not always valid (see
eq.~(\ref{eq:velocitybackward}) below).


\subsection{Limits and special cases}

The discrete-time EQPs have several known models as special cases or
limits.  The following diagram illustrates the relations
between the various models: 
\newcommand{\dst}{\displaystyle}
\begin{eqnarray} \label{eq:relation-queues}
\begin{CD}
 {\rm \fbox{Parallel EQP}} 
 @> \dst \Delta s \to 0 >> 
  {\rm \fbox{Continuous  EQP}} 
 @< \dst \Delta s \to 0 << 
  {\rm \fbox{Backward  EQP}} 
\\
  @V  \dst  p=1    VV  
  @V \dst    p\to\infty   VV
  @V  \dst  p=1    VV  
 \\
{\rm \fbox{
     $\begin{array}{c}
     {\rm Rule~184\ EQP\ with} \\ {\rm stochastic\ boundaries}  
     \end{array}$ 
     }}
   @> \dst \Delta s\to 0 >> 
  {\rm \fbox{Continuous  M/M/1}} 
 @< \dst \Delta s \to 0 << 
  {\rm \fbox{Discrete M/M/1}}.  
\end{CD}
\end{eqnarray}
where $\Delta s$ is the length of the discrete time step.
(A more precise definition of the limit will be given below.) 

\subsubsection{Parallel update with $p=1$}
 
The deterministic hopping cases ($p=1$) of the discrete-time EQPs
correspond to two different processes.  The bulk dynamics of the
parallel EQP with $p=1$ corresponds to the rule 184 cellular automaton
with stochastic boundaries. It is still an EQP although the customer
hopping is deterministic \cite{ref:Y}.  The MC-D and MC-C phases
vanish in the phase diagram (Fig.~\ref{fig:phase-diagrams}).
This case has an exact ``dynamical state'' in a matrix product form
which enables us to derive asymptotic behaviors of the system length,
the number of customers and the density profile in the limit
$t\to\infty$ \cite{ref:AS2}:
\begin{itemize}
\item{Divergent Phase $(\alpha> \beta / (1+\beta))$}: \\
\begin{equation}
\label{eq:184-div}
\langle L_t\rangle =(\alpha-\beta+\alpha\beta)t +o(t) , \quad 
\langle N_t\rangle =\frac{\alpha-\beta+\alpha\beta}{1+\beta}t +o(t) ,
  \quad \rho_{xt,t} \to
   \left\{ \begin{array}{ll} \frac{1}{1+\beta} & (x<1), \\
 0 & (x>1),  \end{array} \right. \qquad 
\end{equation}
\item{Critical Line $(\alpha = \beta / (1+\beta))$:} \\ 
\begin{eqnarray}
\label{eq:184-crit}
\langle L_t\rangle &=& 2\sqrt{\frac{\beta t}{\pi(1+\beta)}} +o(\sqrt{t}) 
\,, \quad 
\langle N_t\rangle =2\sqrt{\frac{\beta t}{\pi(1+\beta)^3}}+o(\sqrt{t}) 
\,, \quad  \nonumber\\
\rho_{x\sqrt{t},t} &\to&  \frac{1}{1+\beta}
{\rm erfc}\left(\frac{x}{2}\sqrt{\frac{1+\beta}{\beta}}\right)\,, 
\end{eqnarray}
\item{Convergent Phase $(\alpha< \beta / (1+\beta))$:}\\ 
\begin{equation}
\label{eq:184-conv}
\langle L_t\rangle \to \frac{\alpha}{\beta-\alpha-\alpha\beta}
\,, \quad 
\langle N_t\rangle \to \frac{\alpha(1-\alpha)}{\beta-\alpha-\alpha\beta}
\,,   \quad 
\rho_{jt} \to  (1-\alpha)\left(\frac{\alpha }{(1-\alpha)\beta}\right)^j,
\end{equation}
where erfc is the complementary error function ${\rm erfc}(x) =
\int^{\infty}_{x}e^{-y^2}dy$.  
\end{itemize}

\subsubsection{Backward-sequential update with $p=1$}

The backward EQP with $p=1$ is equivalent to the discrete-time M/M/1
queueing process which is no longer an EQP as e.g. the state $
\underbrace{1\cdots 1}_{N} $ changes to $ \underbrace{1\cdots 1}_{N-1}
$ when the customer at the server gets service.
Thus no empty site 
appears between the leftmost customer and the server, i.e. we
have always $N_t=L_t$, if the system starts from the empty chain.  
In the limit $t\to\infty$, the system shows different behavior,
depending on the phase:
\begin{itemize}
\item{Divergent Phase $(\alpha>\beta)$:}
\begin{equation}
\langle L_t\rangle = \langle N_t\rangle = (\alpha-\beta)t +o(t)  
\,, \qquad
\rho_{xt,t} \to
\left\{\ \begin{array}{ll} 1 & (x<1), \\ 0 & (x>1),  \end{array} \right.
\label{eq:dmm1-div}
\end{equation}
\item{Critical Line $(\alpha=\beta)$:}\\
\begin{equation}
\langle L_t\rangle = \langle N_t\rangle 
= 2\sqrt{\frac{\beta(1-\beta)}{\pi}t}+o(\sqrt{t})\,, \qquad
\rho_{x\sqrt{t},t} \to {\rm erfc}\left(\frac{x}{2\sqrt{\beta(1-\beta)}}\right),
\label{eq:dmm1-crit}
\end{equation}
\item{Convergent Phase $(\alpha<\beta)$:}
\begin{equation}
\langle L_t\rangle = \langle N_t\rangle \to 
 \frac{\alpha(1-\beta)}{\beta-\alpha}
\,, \qquad
\rho_{jt} \to  \left(\frac{\alpha(1-\beta)}{\beta(1-\alpha)}\right)^j\,.
\label{eq:dmm1-conv}
\end{equation}
\end{itemize}

\subsubsection{Continuous-time update}

Formally, in the continuous-time limit the probabilities $\alpha$,
$\beta$ and $p$ should be replaced by
$  \alpha  \Delta s + o(\Delta s),\ 
     \beta \Delta s + o(\Delta s)$ and 
$  \Delta s + o(\Delta s)$, respectively,  and
time $t$ is rescaled as $t /\Delta s$. 
Then the continuous-time limits $\Delta s\to 0$ of both 
discrete-time EQPs yield the continuous-time EQP studied
in \cite{ref:A}.
The current-density relation for the continuous-time case is simply
\begin{equation}
\label{eq:J-rho-cont}
   J_{\rm cont} (\rho) = p \rho (1-\rho)  ,
\end{equation}
and the phase diagram is given as  
\begin{equation}\quad
   \alpha =  J_{\rm cont}(\rho_{\rm cont}) =
 \left\{ \begin{array}{ll}
     \frac{\beta(p-\beta)}{p}
       & (0<\beta\le p/2 ) ,\\
       \frac{p}{4}
       & (p/2 < \beta < 1) ,
   \end{array}\right.
\quad {\rm where\ \ }
 \rho_{\rm cont}=
 \left\{  \begin{array}{ll}
  1-\frac{\beta}{p}   & ( 0<\beta\le p/2 ), \\
  \frac{1}{2}  & ( \beta > p/2) .
\end{array} \right.
\end{equation}
The continuous-time M/M/1 queueing process is recovered by the
continuous-time limits of the rule 184 case and the discrete-time M/M/1
queue.  It is also obtained by  the $p\to\infty$ limit of the
continuous-time EQP.


\section{Subphases in the divergent phase}\label{sec:subphases}

\begin{figure}
\begin{center}
\vspace{-5mm}
\includegraphics[width=0.375\columnwidth]{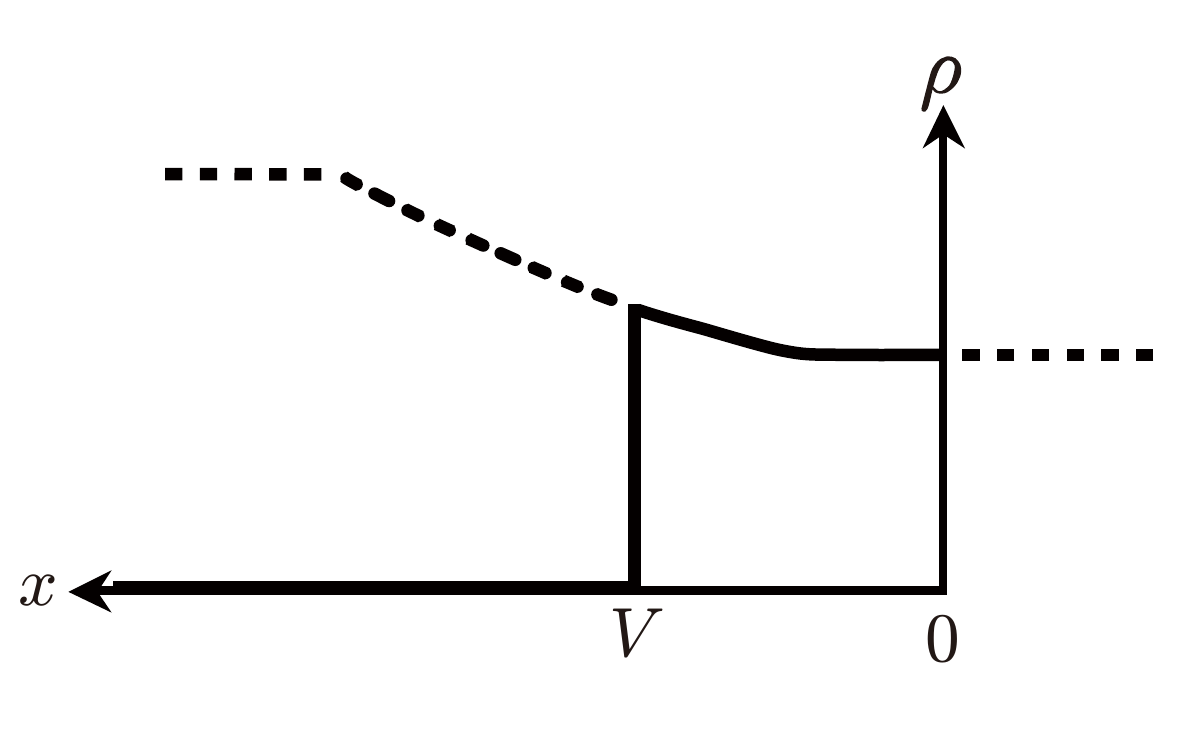}
\vspace{-5mm}
\caption{A schematic picture for the density profile
  in the divergent phase, where $x$ is the rescaled position $j/t$.
  According to the injection probability (rate) $\alpha$, the
  rarefaction wave is ``cut'' by the leftmost customer ($x=V$) and the
  server ($x=0$).  }
\label{fig:rarefaction}
\end{center}
\end{figure}

We consider the TASEP on an infinite chain with the initial
densities $\rho_{\rm right}$ (at  sites $j<0$) 
 and $\rho_{\rm left}$ (at  sites $j\ge 0$), where
$\rho_{\rm left}>\rho_{\rm right}$.  When the current $J$ from left to
right is given by a function of the density $\rho$,
 the rescaled density profile $\rho(x=j/t)$ is well-described by
\begin{equation}
\label{eq:rho(x)}
  \rho (x) \simeq  \left\{ \begin{array}{ll}
   \rho_{\rm right}  & (x<f(\rho_{\rm right})  ),  \\
   f^{-1}(x)  & (f(\rho_{\rm left})>x>f(\rho_{\rm right}) ), \\
   \rho_{\rm left}  & (x>f(\rho_{\rm left})  ) 
  \end{array} \right.
\end{equation}
with $f(\rho) = -\frac{dJ}{d \rho}$ \cite{ref:KRB}.  We will see that,
for the EQPs, the density profiles in the divergent phase 
are obtained by cutting this
``rarefaction wave'' as in Fig.~\ref{fig:rarefaction}.

\subsection{Parallel case}

From the current-density relation (\ref{eq:J-rho-para}) for the
parallel-update TASEP we have
\begin{equation}
f_{\sss \parallel} (\rho)
 =-\frac{dJ_{\sss\parallel}}{d\rho} =
\frac{p(2\rho -1)}{\sqrt{1-4p\rho(1-\rho)} }\, ,
\quad 
f_{\sss \parallel}^{-1} (x)
 = \frac{1}{2} +  \frac{x}{2} \sqrt{\frac{1-p}{p(p-x^2)}  }\,.
\end{equation}
We assume that the (rescaled) density profile $\rho_{xt,t}$
(at site $xt$ and time $t$) has the form
\begin{equation}
\label{eq:rho_xtt} 
  \rho_{xt,t} \simeq
    \left\{\begin{array}{ll}
   0         & (x>V,0>x),  \\
   \rho (x)  & (V>x>0), 
  \end{array}\right.
\end{equation}
where $\rho(x)$ is given by (\ref{eq:rho(x)})
with $\rho_{\rm right}=\rho_{\sss\parallel}$ and $\rho_{\rm left}=1$.
This assumption is supported by simulation results.
Here $V$ is the velocity of the system length 
 $ \langle L_t\rangle \simeq Vt $.

\begin{figure}[h]
\begin{center}
  \includegraphics[width=0.32\columnwidth]{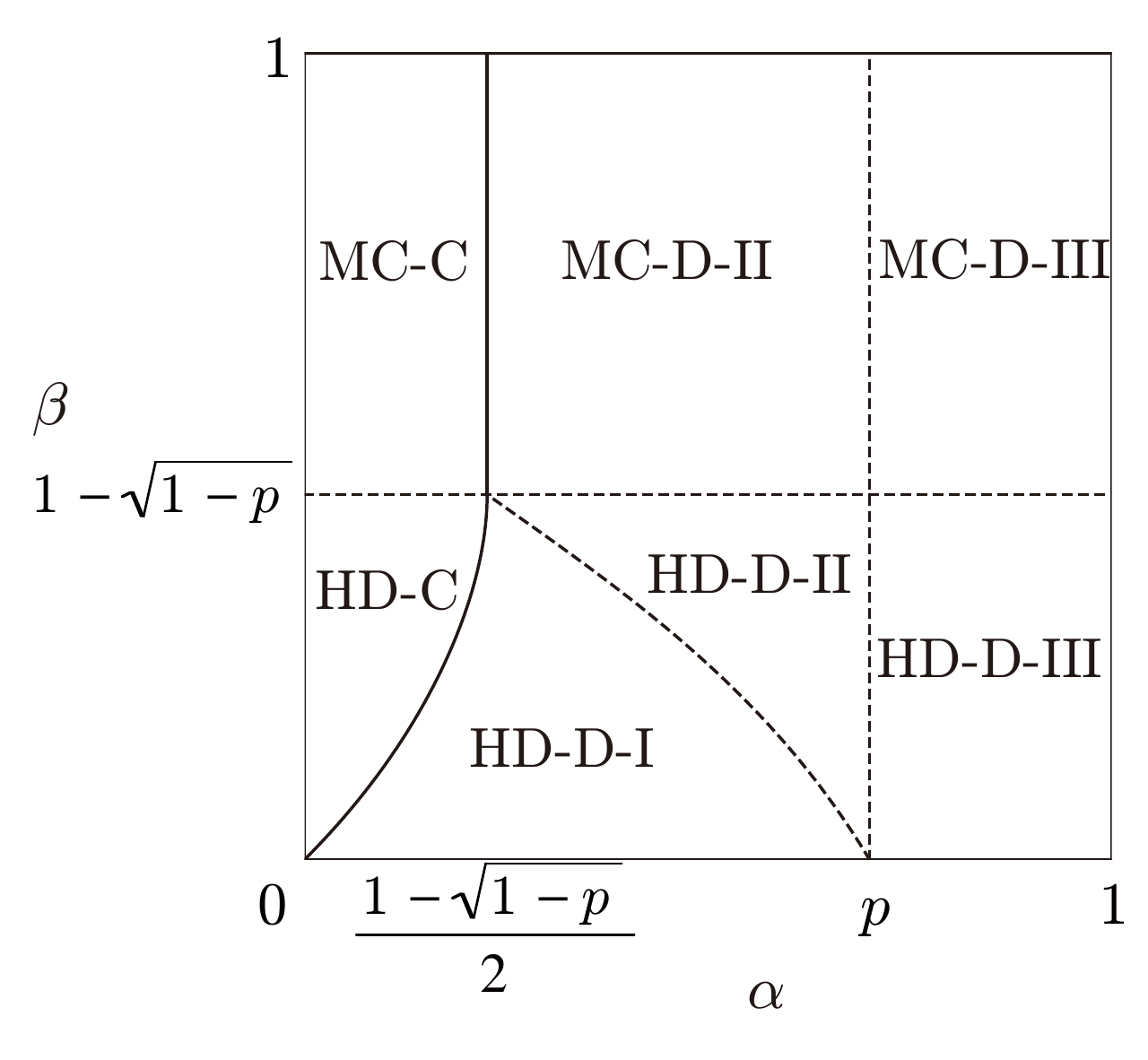}\qquad
 \includegraphics[width=0.32\columnwidth]{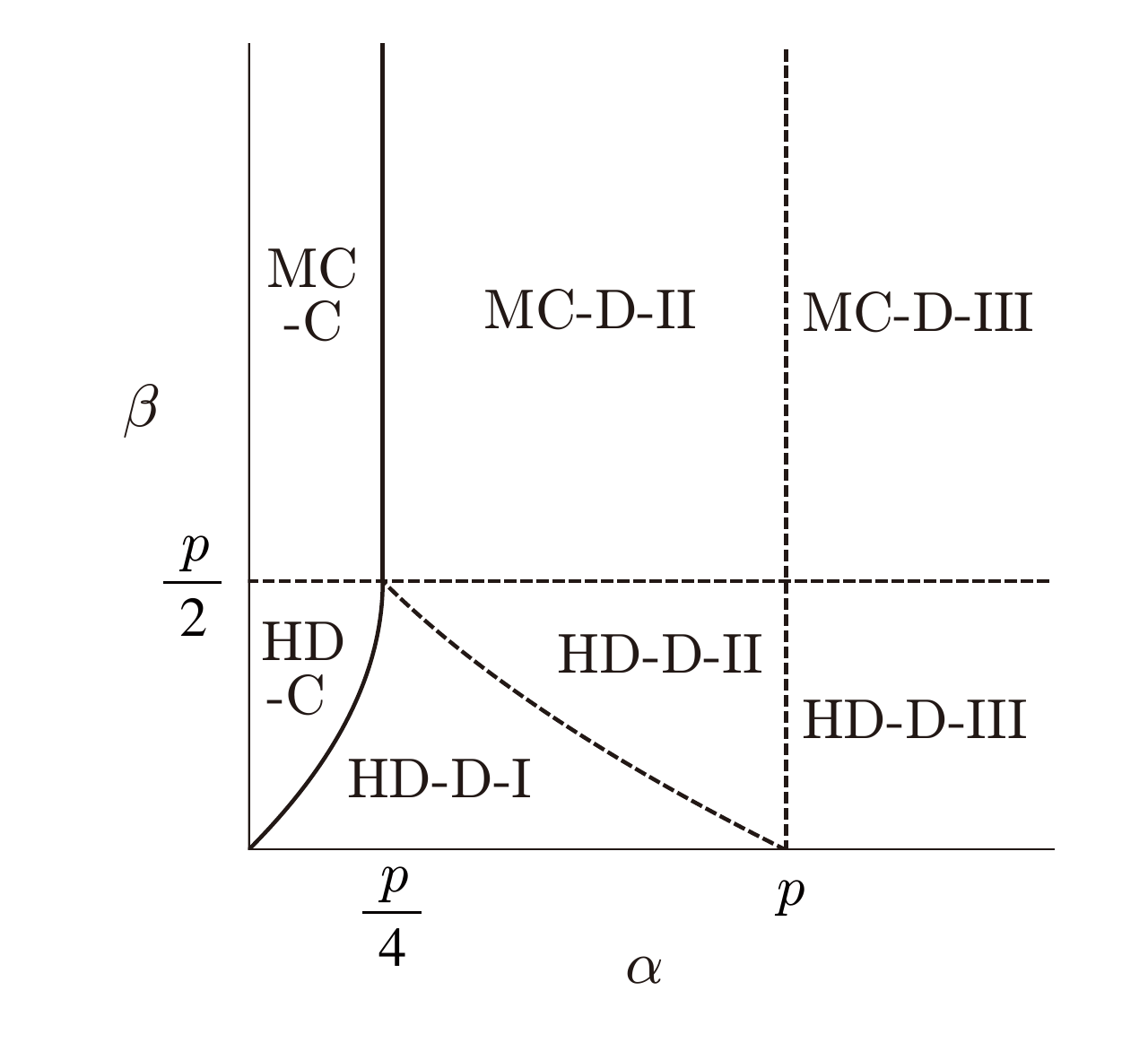}
 \\
 \includegraphics[width=0.32\columnwidth]{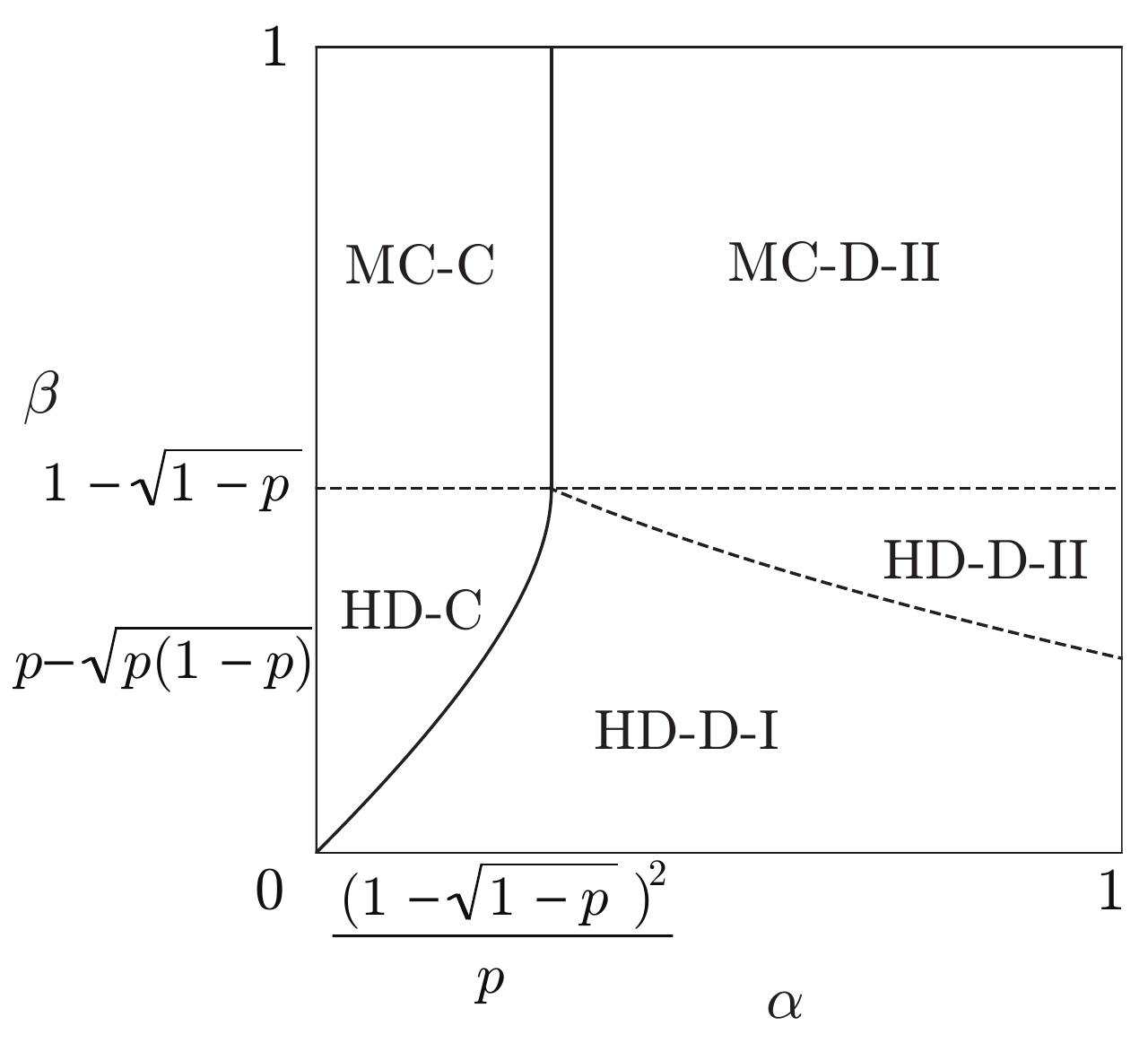}\qquad
 \includegraphics[width=0.32\columnwidth]{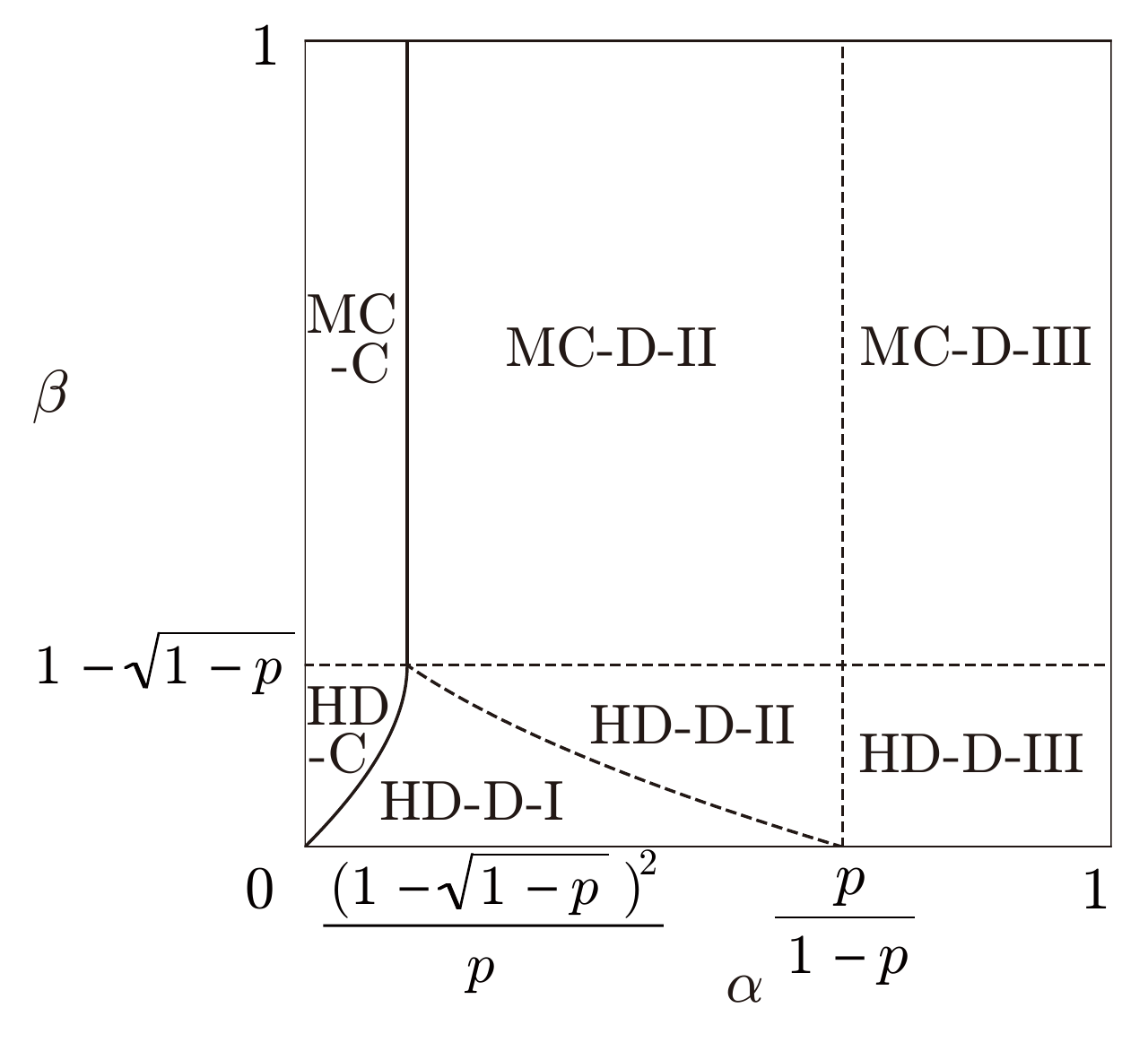}
\vspace{-5mm}
\caption{
The subphases of the EQPs;
the parallel EQP (top-left),
the continuous-time EQP (top-right),
the backward EQP with $\frac{1}{2}\le p<1$ (bottom-left) and 
the backward EQP with $0<p<\frac{1}{2}$ (bottom-right).
}
\label{fig:pd}
\end{center}
\end{figure}

\begin{figure}
\begin{center}
 \includegraphics[width=0.4\columnwidth]{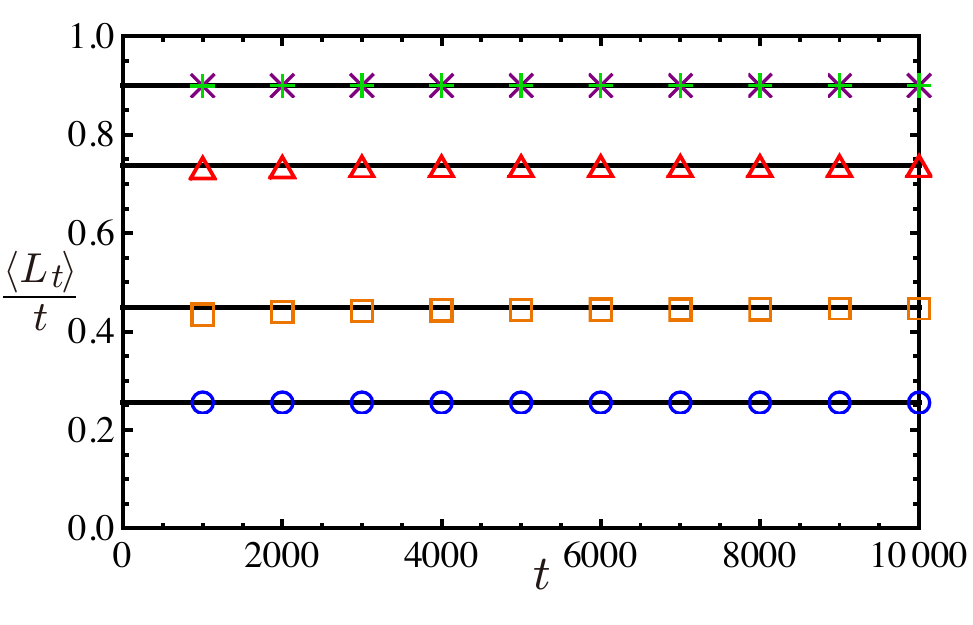}
 \includegraphics[width=0.4\columnwidth]{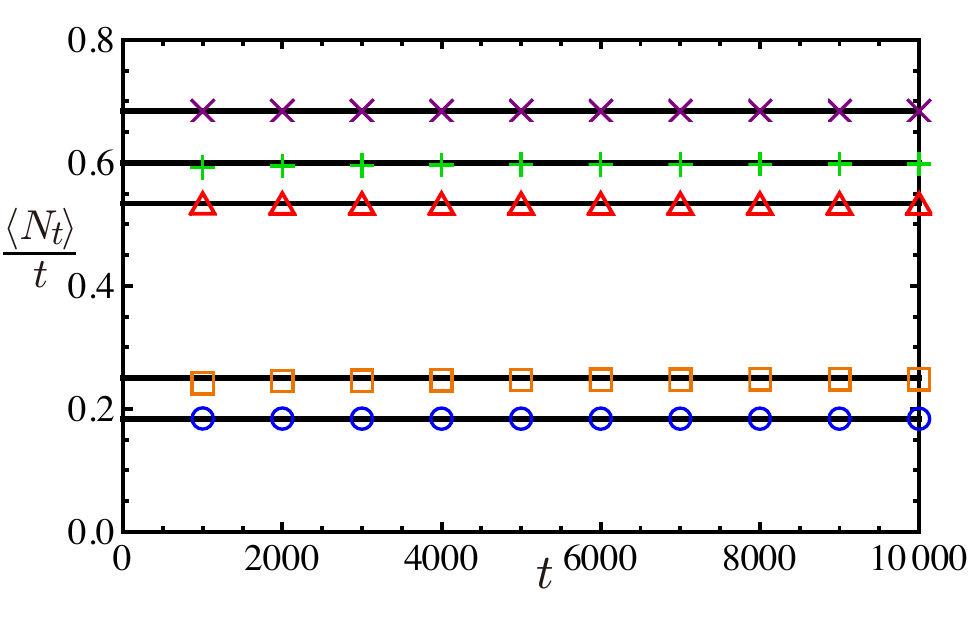}
\\
\vspace{-2mm}
 \includegraphics[width=0.4\columnwidth]{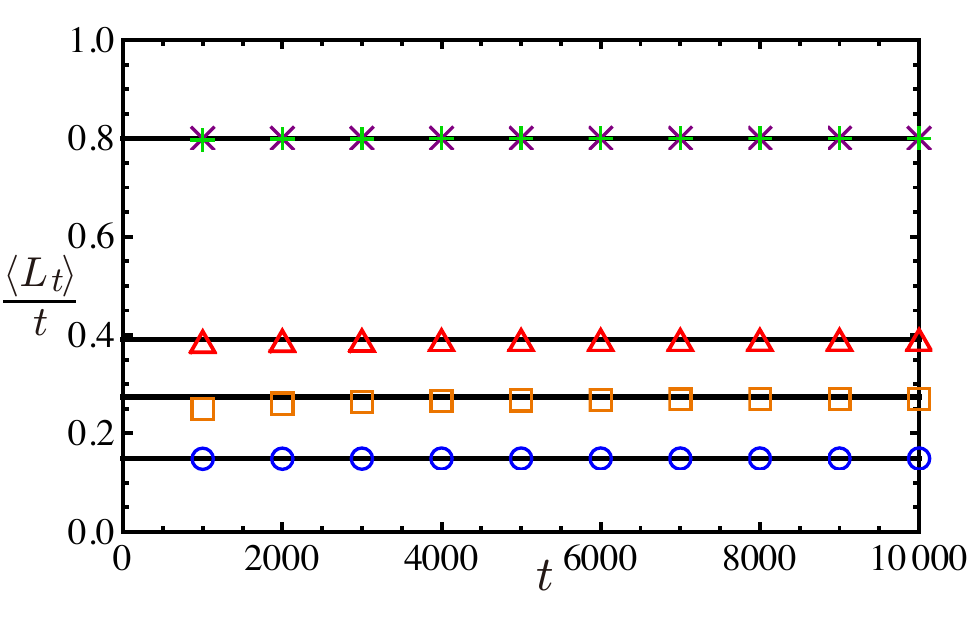}
 \includegraphics[width=0.4\columnwidth]{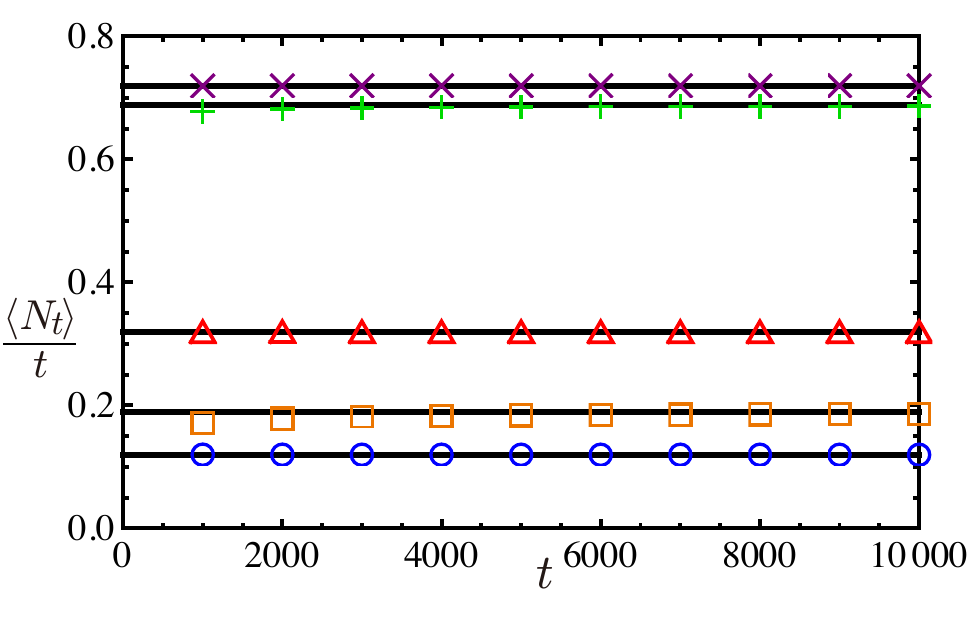}
\\
\vspace{-2mm}
 \includegraphics[width=0.4\columnwidth]{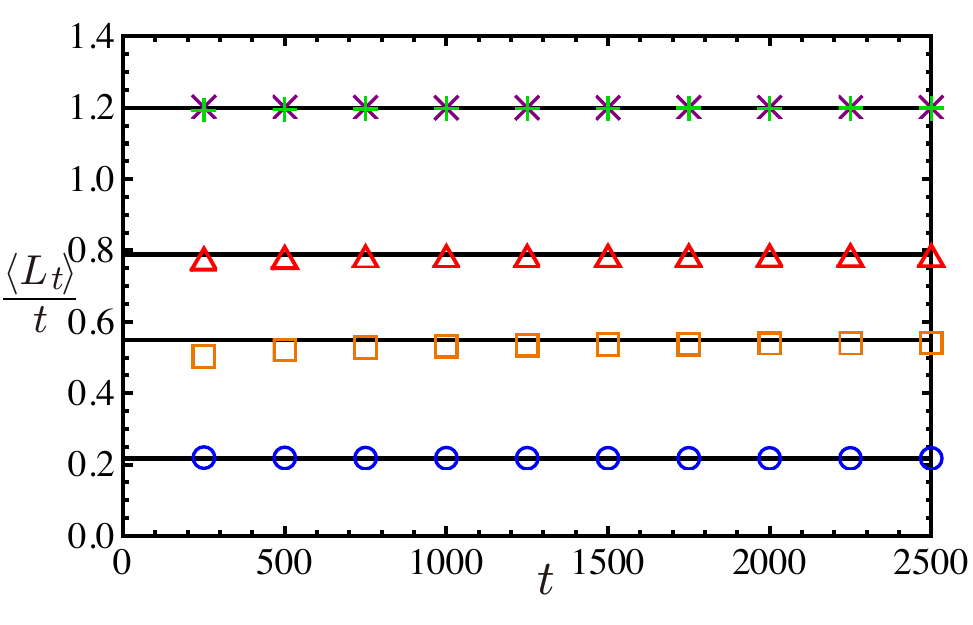}
 \includegraphics[width=0.4\columnwidth]{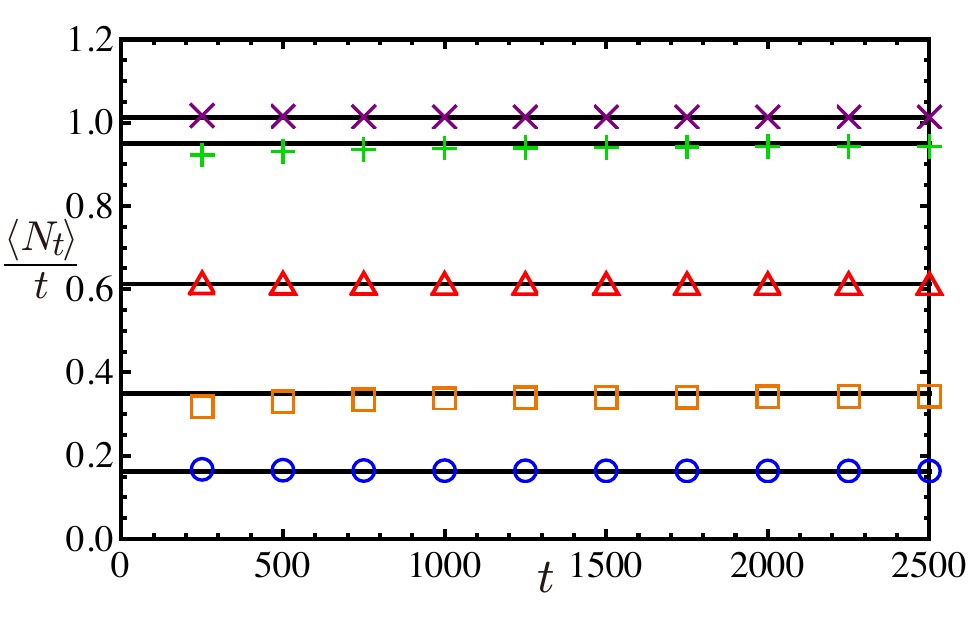}
\caption{
  The  growth  velocities of the system size and the number of
  customers for the parallel (top), backward (middle) and continuous
  (bottom) EQPs.  The simulation data were obtained by averaging $10^4$
  samples.  We see that these agree with the lines corresponding to
  (\ref{eq:velo-L-para}). The parameters are chosen as in
  Equations (\ref{eq:parameter-para}), (\ref{eq:parameter-back}) and
  (\ref{eq:parameter-cont}).   }
\label{fig:velocities}
\end{center}
\end{figure}

\begin{figure}
\begin{center}
 \includegraphics[width=0.4\columnwidth]{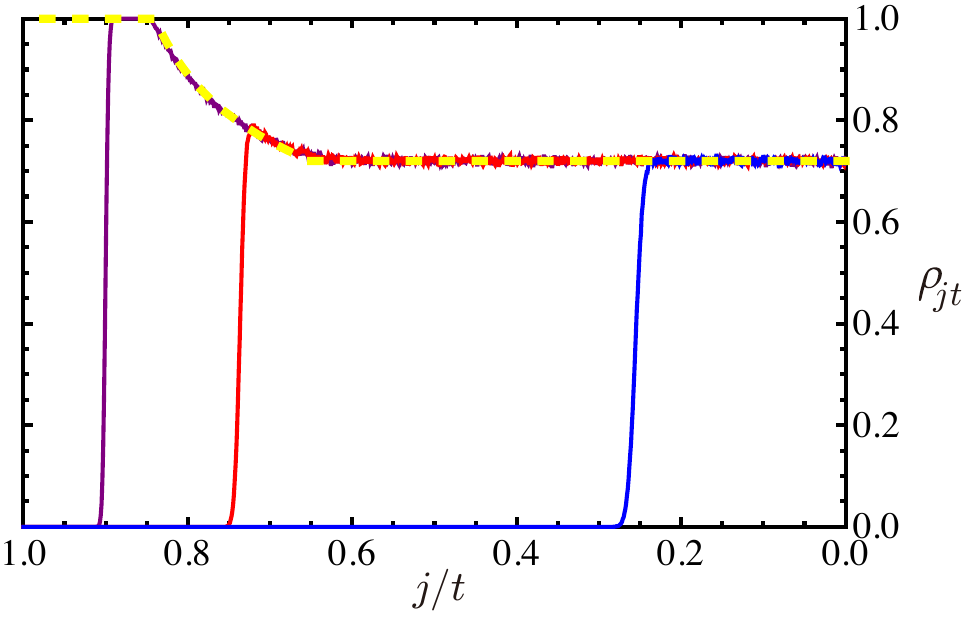}\qquad
 \includegraphics[width=0.4\columnwidth]{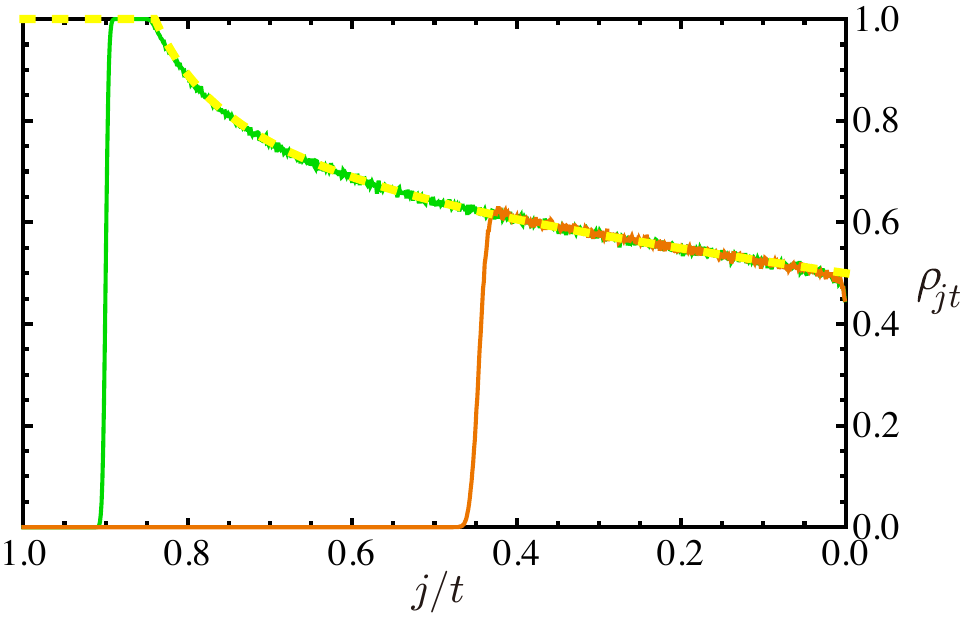}
\\
 \includegraphics[width=0.4\columnwidth]{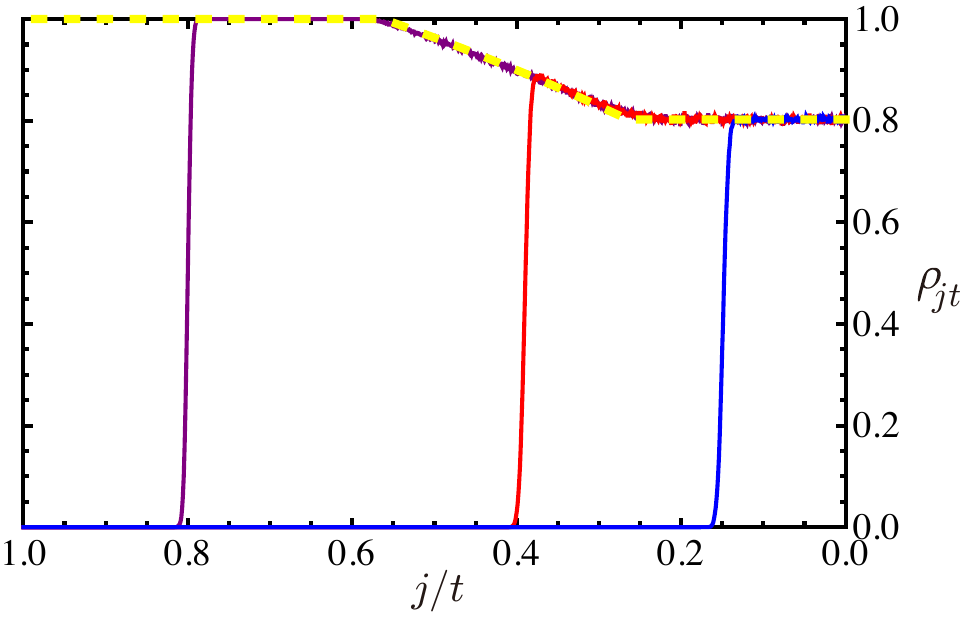}\qquad
 \includegraphics[width=0.4\columnwidth]{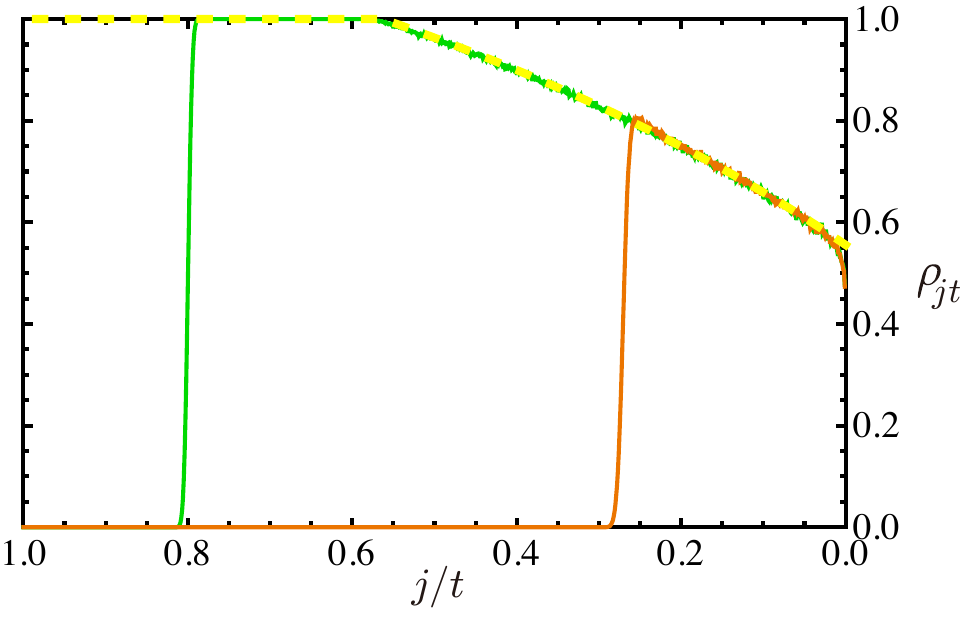}
\\
 \includegraphics[width=0.4\columnwidth]{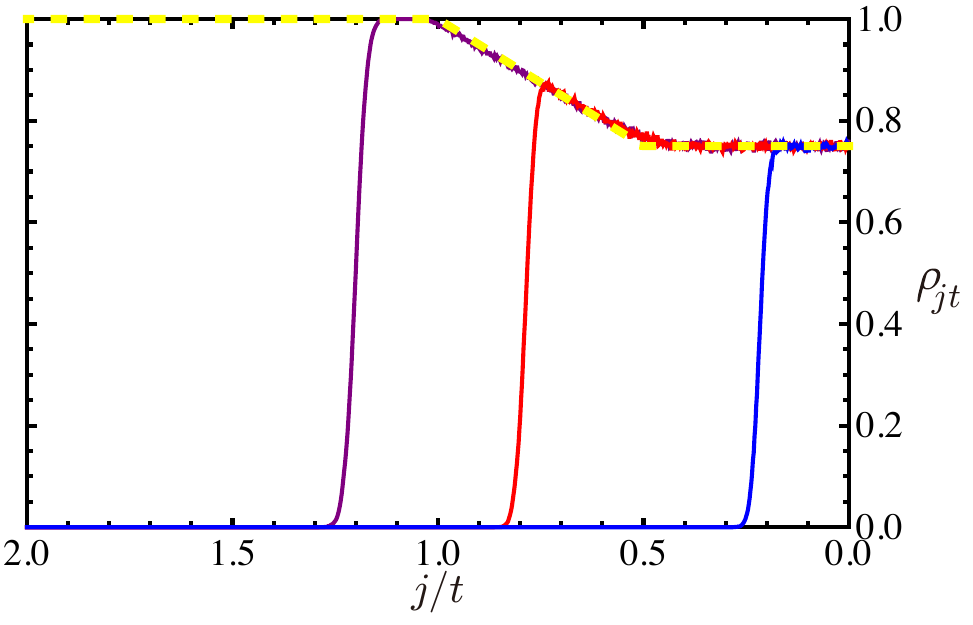}\qquad
 \includegraphics[width=0.4\columnwidth]{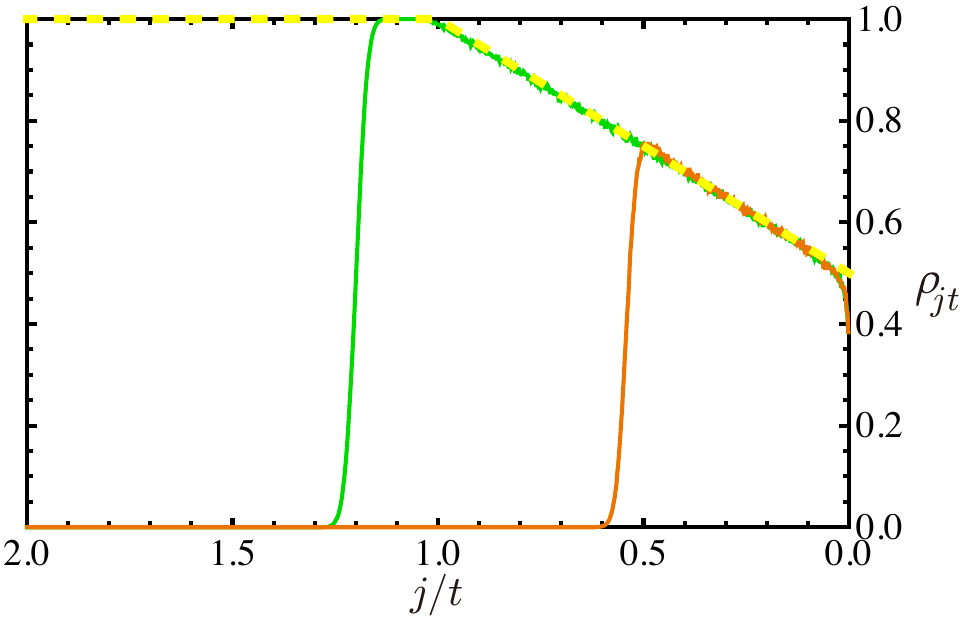}
\\
\caption{
  Rescaled density profiles $\rho_{jt}$ of the parallel
  (top, $t=10^4 $ ), backward (middle,  $t=10^4 $)
   and continuous-time (bottom, $t=2500$)  EQPs.  The
  parameters are chosen as in Equations (\ref{eq:parameter-para}),
  (\ref{eq:parameter-back}) and (\ref{eq:parameter-cont}).  The
  simulation data were obtained by averaging $10^4$ samples. }
  \label{fig:div-density-profiles}
\end{center}
\end{figure}

Under the assumption (\ref{eq:rho_xtt}) we have 
\begin{equation}
\label{eq:N=integral}
  t(\alpha - J^{\rm out}  )
  \simeq t \int_0^V \rho(x) dx ,
\end{equation}
where both sides are different expressions for the number of
customers. 
 Inserting $J^{\rm out} = J_{\sss\parallel}
(\rho_{\sss\parallel})$ (see Equation (\ref{eq:crit-line-para})) into 
Equation (\ref{eq:N=integral}), we find the velocity 
\begin{equation}\label{eq:velo-L-para}
  V=V_{\sss \parallel} =
  \left\{\ \begin{array}{ll}
  \frac{\alpha-J(\rho_{\sss \parallel}) }
  {\rho_{\sss \parallel} } = 
       \alpha\frac{p-\beta^2}{p-\beta} - \beta
        & ({\rm I}), \\
       2p\alpha-p+2\sqrt{p\alpha(1-p)(1-\alpha)}
        & ({\rm II}), \\
       \alpha & ({\rm III}),
  \end{array}\right.
\end{equation}
where 
\begin{eqnarray}
\begin{array}{clcc}
{\rm I}:&\ 0<V_{\sss \parallel} \le 
   f_{\sss\parallel} (\rho_{\sss\parallel})
   \quad \phantom{f_{\sss\parallel}(1)} {\rm i.e.} \quad
   \frac{\beta(p-\beta)}{p-\beta^2}
   <\alpha \le \frac{(p-\beta)^2}{p-2p\beta+\beta^2},\\ 
{\rm II}:&\  
   f_{\sss\parallel}(\rho_{\sss\parallel})
   <  V_{\sss \parallel} \le  f_{\sss\parallel}(1)
   \quad \phantom{0} {\rm i.e.} \quad
{\rm Max}  \left(  \frac{(p-\beta)^2}{p-2p\beta+\beta^2},
 \frac{1-\sqrt{1-p}}{2}  \right) 
 < \alpha \le p,\\
{\rm III}:&\  
  f_{\sss\parallel} (1) \ge  V_{\sss \parallel} 
   \quad \phantom{f_{\sss\parallel}(\rho_{\sss\parallel})<0}{\rm i.e.} \quad
 p < \alpha \le 1 .
\end{array}
\end{eqnarray}
Combing this and the form of $\rho_{\sss\parallel}$  given in
Equation (\ref{eq:crit-line-para}),
we obtain five subphases in the divergent
phase.  In each phase the rescaled density $\rho_{xt,t}$ has a
different form (Fig.~\ref{fig:pd}):
\begin{eqnarray}
\label{eq:dp}
\begin{array}{l}
$HD-D-I$:
\rho_{xt,t}\simeq
\left\{\ \begin{array}{ll}
   \rho_{\rm right} & (V>x>0), \\
  0 & (x>V),
\end{array}\right.  \\
$HD-D-II$:\   
\rho_{xt,t}\simeq
\left\{ \begin{array}{ll}
   \rho_{\rm right} & (v_1>x>0), \\
   f^{-1} (x) & (V>x>v_1), \\
  0 & (x>V),
\end{array} \right. \ 
$MC-D-II$:\   
\rho_{xt,t}\simeq
\left\{ \begin{array}{ll}
   f^{-1} (x) & (V>x>0), \\
  0 & (x>V),
\end{array} \right.  \\
$HD-D-III$:\   
\rho_{xt,t}\simeq
\left\{ \begin{array}{ll}
   \rho_{\rm right} & (v_1>x>0), \\
   f^{-1} (x) & (v_2>x>v_1), \\
   1 & (V>x>v_2), \\
  0 & (x>V),
\end{array} \right. \ 
$MC-D-III$:\   
\rho_{xt,t}\simeq
\left\{ \begin{array}{ll}
   f^{-1} (x) & (v_2>x>0), \\
   1 & (V>x>v_2), \\
  0 & (x>V),
\end{array}\right. 
\end{array}
\end{eqnarray}
where
\begin{equation}\quad
\rho_{\rm right}=\rho_{\sss\parallel}\,,\quad
v_1= f_{\sss\parallel}  (\rho_{\sss\parallel})
  = \frac{p(p-2\beta+\beta^2)}{p-2p\beta+\beta^2}\,,\quad 
v_2=f_{\sss\parallel}(1)=p\,,\quad 
f^{-1}(x)=f^{-1}_{\sss\parallel}(x)\,.
\end{equation}
The HD-D phase is divided into three phases: (I) plateau, (II)
plateau-slope and (III) plateau-slope-plateau.  On the other hand, the
MC-D phase is divided into only two phases: (II) slope and (III)
slope-plateau. The plateau near the exit does not appear.
Figures \ref{fig:velocities} and \ref{fig:div-density-profiles} show
simulation results for the velocities and the density profiles,
respectively, with parameters
\begin{eqnarray}\label{eq:parameter-para}
(\alpha,\beta,p) =  
\left\{ \begin{array}{lll}
(0.4,0.3,0.84) & $HD-D-I$   &  {\scriptstyle\bigcirc}\  $(blue)$, \\
(0.75,0.3,0.84) & $HD-D-I$ &  \triangle\ $(red)$,  \\
(0.9,0.3,0.84) & $HD-D-III$ &  \times\   $(purple)$, \\
(0.55,0.8,0.84) & $MC-D-II$  &   \square\  $(orange)$, \\
(0.9,0.8,0.84) & $MC-D-III$  & +   \    $(green)$.   
 \end{array}\right.
\end{eqnarray}

\subsection{Backward case}

From the current-density relation (\ref{eq:J-rho-back}) for the TASEP
with the backward update we have
\begin{equation}
   f_{\sss \gets}(x) := -\frac{d J_{\sss\gets}}{d\rho}
   =-\frac{p(1-2\rho+p\rho^2)}{(1-p\rho)^2}\, ,\qquad
   f^{-1}_{\sss\gets} (x)
   = \frac{1}{p} - \frac{1}{p}
    \sqrt{ \frac{ 1-p}{ 1+x} } \,.
\end{equation}
As in the parallel-update case, we assume Equation (\ref{eq:rho_xtt})
with $\rho(x)$ as in (\ref{eq:rho(x)}), $\rho_{\rm right}=\rho_{\sss\gets}$
and $\rho_{\rm left}=1$.  From Equation (\ref{eq:N=integral}), we find
that the velocity $V_{\sss\gets}$ for the system length $\langle L_t
\rangle \simeq V_{\sss\gets} t$ is given by
\begin{equation}
  V_{\sss \gets} =
  \left\{ \begin{array}{ll}
  \frac{\alpha-J(\rho_{\sss \gets}) }
  {\rho_{\sss \gets} } = 
  \frac{p(1-\beta)}{p-\beta}\alpha-\beta
        & \qquad ({\rm I}), \\
       2\sqrt{p (1-p)\alpha -p(1-\alpha)}
        & \qquad ({\rm II}), \\
       \alpha & \qquad ({\rm III}),
  \end{array} \right.
\label{eq:velocitybackward}
\end{equation}
where 
\begin{eqnarray}
\begin{array}{clcc}
{\rm I}:&\  0<V_{\sss\gets} \le 
   f_{\sss\gets} (\rho_{\sss\gets})
   \quad \phantom{f_{\sss\gets}(1)}{\rm i.e.} \quad  
     \frac{\beta(p-\beta)}{p(1-\beta)}
   <\alpha \le \frac{(p-\beta)^2}{p(1-p)} ,\\ 
{\rm II}:&\    
   f_{\sss\gets}(\rho_{\sss\gets})
   <V_{\sss\gets} \le f_{\sss\gets}(1)
  \quad \phantom{0}{\rm i.e.} \quad 
{\rm Max}  \left(  \frac{(p-\beta)^2}{p(1-p)} ,
 \frac{(1-\sqrt{1-p})^2  }{p}  \right) 
 < \alpha \le \frac{p}{1-p},\\
{\rm III}:&\    
  f_{\sss\gets} (1) > V_{\sss\gets}
   \quad \phantom{f_{\sss\gets}(\rho_{\sss\gets})<0}{\rm i.e.} 
   \quad  \frac{p}{1-p} < \alpha \le 1 \,.
\end{array}
\end{eqnarray}
When $\frac{1}{2}\le p<1$, the case D-III vanishes and the divergent
phase is divided into three phases (Fig.~\ref{fig:pd}).  On the
other hand, when $0<p<\frac{1}{2}$, the structure of the subphases is
qualitatively similar to the parallel case.  The density profiles are
given by Equation (\ref{eq:dp}) with
\begin{equation}\quad
\rho_{\rm right}= \rho_{\sss\gets},\quad 
v_1= f_{\sss\gets}  (\rho_{\sss\gets})
  = \frac{p-2\beta+\beta^2 }{ 1-p},\quad  
v_2= f_{\sss\gets}(1)=\frac{p}{1-p},\quad 
f^{-1}(x)=
f^{-1}_{\sss\gets} (x).
\end{equation}
Figures~\ref{fig:velocities} and \ref{fig:div-density-profiles} show
simulation results of the velocities and the density profiles,
respectively, with parameters
\begin{eqnarray}\label{eq:parameter-back}
(\alpha,\beta,p) =  
\left\{ \begin{array}{lll}
 (0.2,0.1,0.36) & $HD-D-I$   &  {\scriptstyle\bigcirc} \  $(blue)$, \\
 (0.4,0.1,0.36) & $HD-D-I$ &  \triangle\ $(red)$,  \\
 (0.8,0.1,0.36) & $HD-D-III$ &  \times\   $(purple)$, \\
 (0.3,0.6,0.36) & $MC-D-II$  &   \square\  $(orange)$, \\
 (0.8,0.6,0.36) & $MC-D-III$  & +   \    $(green)$.   
 \end{array}\right.
\end{eqnarray}

\subsection{Continuous-time case}

From the current-density relation (\ref{eq:J-rho-cont}) for the
continuous-time TASEP, we have
\begin{eqnarray}
f_{\rm cont} (\rho)
:=-\frac{dJ_{\rm cont}}{d\rho} =p(2\rho -1) .
\end{eqnarray}
The velocity $V_{\rm cont}$ of the system length, the subphases and
the density profiles can be obtained following the same procedure as for the
parallel and backward EQPs, or simply by taking the continuous-time
limits of the results for the two discrete cases:
\begin{eqnarray}
  V_{\rm cont} =
\left\{  \begin{array}{ll}
  \frac{\alpha-J(\rho_{\rm cont}) }
  {\rho_{\rm cont} } = 
  \frac{p}{p-\beta}\alpha-\beta   & ({\rm I}), \\
       2\sqrt{p \alpha } -p & ({\rm II}), \\
       \alpha & ({\rm III}),
  \end{array}\right.
\end{eqnarray}
where 
\begin{eqnarray}
{\rm I}:&\  0<V_{\rm cont} \le 
   f_{\rm cont}(\rho_{\rm cont})
   \quad \phantom{f_{\rm cont}(1)}{\rm i.e.} \quad  \frac{\beta(p-\beta)}{p}
   <\alpha \le \frac{(p-\beta)^2}{p} ,\\ 
{\rm II}:&\    
   f_{\rm cont} (\rho_{\rm cont})
   <V_{\rm cont} \le f_{\rm cont}(1)
  \quad \phantom{0}{\rm i.e.} \quad 
{\rm Max}  \left(  \frac{(p-\beta)^2}{p} ,
 \frac{p}{4}  \right) 
 < \alpha \le p ,\\
{\rm III}:&\    
  f_{\rm cont} (1) > V_{\rm cont}
   \quad \phantom{f_{\rm cont}(\rho_{\rm cont})<0}{\rm i.e.} 
   \quad p < \alpha \le 1 .
\end{eqnarray}

The density profiles are given by Equation  (\ref{eq:dp}) with
\begin{eqnarray}
\rho_{\rm right}= \rho_{\rm cont},\ 
v_1= f_{\rm cont}  (\rho_{\rm cont})  = p-2\beta  ,\ 
v_2= f_{\rm cont}(1)=p,\ 
f^{-1}(x)=f^{-1}_{\rm cont} (x) = \frac{1}{2} + \frac{x}{2p}  .
\end{eqnarray}
Figures \ref{fig:velocities} and \ref{fig:div-density-profiles}
 show  simulation results 
of the velocities and the density profiles, respectively,
 with parameters
 
 \begin{eqnarray}\label{eq:parameter-cont}
(\alpha,\beta,p) =  
\left\{ \begin{array}{lll}
 (0.35,0.25,1)  & $HD-D-I$   &  {\scriptstyle\bigcirc}\  $(blue)$, \\
 (0.8,0.25,1) & $HD-D-I$ &  \triangle\ $(red)$,  \\
 (1.2,0.25,1) & $HD-D-III$ &  \times\   $(purple)$, \\
 (0.6,1, 1) & $MC-D-II$  &   \square\  $(orange)$, \\
 (1.2,1,1) & $MC-D-III$  & +   \    $(green)$.   
 \end{array}\right.
\end{eqnarray}


\section{On the critical line}\label{sec:critical}

Since $\langle N_t\rangle$ and $\langle L_t\rangle$ converge to
stationary values in the convergent phase, and diverge proportional to
$t$ in the divergent phase, 
we expect that  they behave as 
\begin{equation}
\langle X_t\rangle \sim t^{\gamma_X}, \qquad
 0\le  \gamma_X \le 1,\qquad {\rm for} \ X=L,N\, .
\label{eq:powerlaw}
\end{equation}

\begin{figure}[h]
\begin{center}
 \includegraphics[width=0.42\columnwidth]{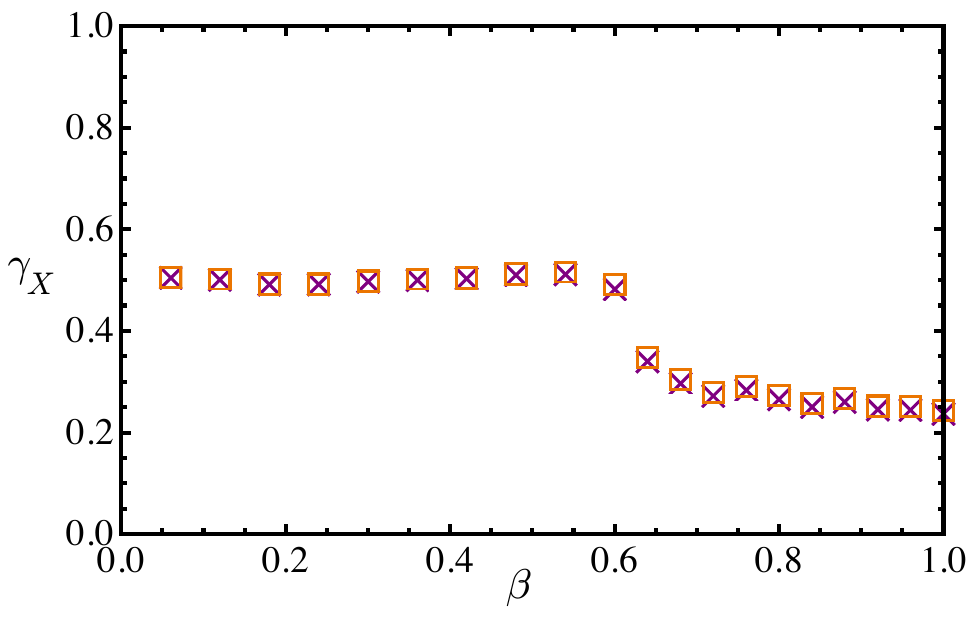}\qquad
 \includegraphics[width=0.42\columnwidth]{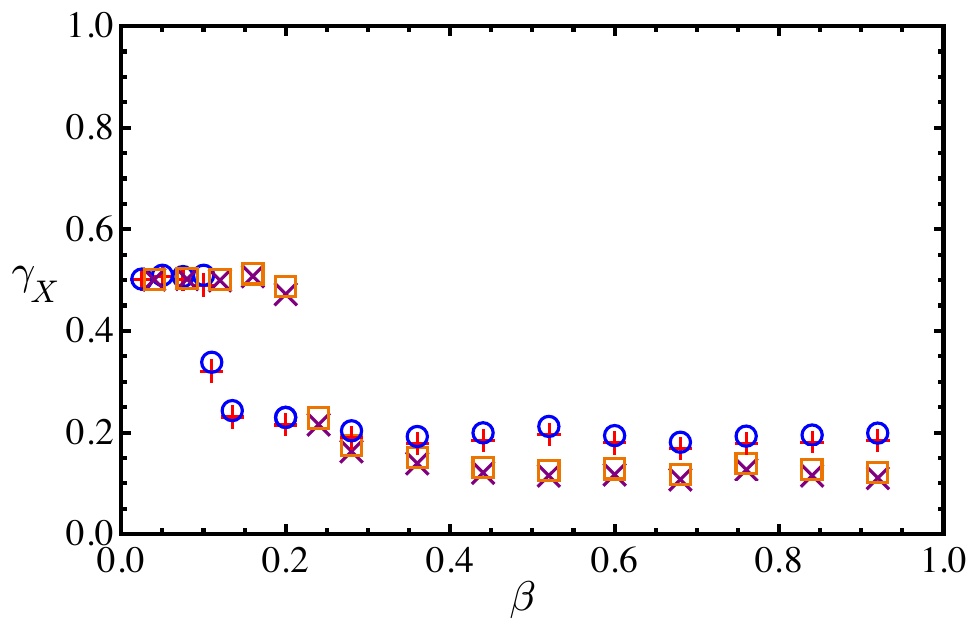}
\vspace{-5mm}
\caption{The exponents $\gamma_X$ for
 the system length $X=L$ and the number of particles $N$
 on the critical lines of the
  parallel ($p=0.84,\ \beta_c=0.6$, left) and backward ($p=0.36,\ 
  \beta_c=0.2$ and $p=0.19,\  \beta_c=0.1$, right) EQPs.  
 The markers $\square$ (orange) and $\times$ (purple)   
 correspond  to $\gamma_L$ and $\gamma_N$,  respectively, 
 for the parallel case with $p=0.84$ and the backward case  with $p=0.36$,
 and  $\scriptstyle\bigcirc$ (blue) and   $+$ (red)  correspond to 
  $\gamma_L$ and $\gamma_N$,  respectively, for  the backward case with $p=0.19$.
 }
\label{fig:LN-on-critical}
\end{center}
\end{figure}

Under this assumption we have 
\begin{equation}
 \frac{\ln X_t - \ln X_{t/b}}{\ln b}
 \to \gamma_X\qquad  (t\to\infty)\,,
\label{eq:exponents}
\end{equation}
for $X=L,N$.  After verifying that the growth behavior is indeed
well-described by power-laws of the form (\ref{eq:powerlaw}), we
estimate the exponents $\gamma_X$ by applying (\ref{eq:exponents}) to
 simulated samples with $b=10$ and $t\le 5\times 10^5$.
The number of samples for this estimation for each parameter set
is basically  $10^4$,  but $10^6$ or $5\times 10^6$ samples were used 
for the backward EQP  in the region  $0.28\le \beta< 0.8$ and 
$ \beta> 0.8$, respectively, because  there  fluctuations of $L$ and $N$ 
are very large.
The results shown in Fig.~\ref{fig:LN-on-critical} are consistent
with the expectation $\gamma_L=\gamma_N$ everywhere on the critical
line.  This is supported by the observation that the total density
$\rho_{\rm tot}=\langle N_t \rangle/\langle L_t \rangle$ reaches
quickly an almost stationary value which implies that
$\gamma_L=\gamma_N$.  More detailed results will be
present in a future publication.

The critical lines of the EQPs consist of two parts: a curved and a
straight line (Fig.~\ref{fig:phase-diagrams}).
On the curved part, the simulation results indicate
\begin{equation}
 \langle N_t\rangle,\langle L_t\rangle =O\left(\sqrt{t}\right)\,.  
\end{equation}

The behavior of $\langle L_t\rangle$ and $\langle N_t\rangle$ on the
straight part of the critical line is not so clear although diffusive
behavior can be excluded.  As Fig.~\ref{fig:LN-on-critical} indicates,
the exponents are smaller than on the curved part, i.e.
$\gamma_L=\gamma_N<1/2$.  For the parallel case, $\gamma =1/4$ (with
large corrections near $\beta_c$) can not be excluded, but for the
backward case, the exponents seems to depend on the value of $p$.
For example, the exponents for $p=0.19$ seem to be bigger than those
for $p=0.36$, see the right graph of Fig.~\ref{fig:LN-on-critical}. 
  Our simulation
results are not sufficient to determine conclusively the
dependency of $\gamma$ on the parameters, e.g. how it varies with $\beta$
near $\beta_c$.


\section{Conclusion}\label{sec:conclusion}

We have continued our studies of the exclusive queueing process
(EQP) which extends the classical M/M/1 queueing process by
incorporating the  excluded volume effect.
We have compared the behavior of the model with different
update schemes (parallel, backward-sequential, continuous time).
The phase diagrams are qualitatively similar, except for certain
limiting cases.

The phase diagram of the EQP turns out to be rather rich.  Here we
have shown that the divergent phase is subdivided into up to 5
different subphases according to the parameter dependence of the
current and the density profiles.
The MC-D phase has two
different subphases (the slope and plateau-slope phases),
and the HD-D phase has three different subphases (the plateau,
plateau-slope and plateau-slope-plateau phases).

In the divergent phase we have conjectured
the analytic form of the density profiles which show good agreement
with simulation results. 
The shapes of the rescaled profiles can be understood in terms of a rarefaction
wave that is ``cut'' at both ends.

On the critical line separating the divergent from the convergent
phase the length of the system grows sublinearly. Based on simulation
results we find  diffusive behavior on the
curved part of the critical line (i.e.\ $\beta < \beta_c$) for all
updates. 
In the special case $p=1$ for the two discrete EQPs, 
the density profiles can be written in terms of the complementary 
error function as Equations  (\ref{eq:184-crit}) and (\ref{eq:dmm1-crit}). 
Identifying the density profile on the curved part for the EQPs 
with general values of $p$ is one of problems that  need
to be clarified in the future.

The behavior on the straight part $\beta > \beta_c$ of the
critical line is subdiffusive ($\gamma <1/2$).  However we could not
clearly determine whether the exponent $\gamma$ depends on the
parameters, and more simulation data with sufficient accuracy are
needed to determine the behavior on the straight part.


\section*{Acknowledgement}
C Arita is a JSPS fellow for research abroad.
The authors thank Kirone Mallick for useful discussions.


\end{document}